\documentclass[%
 aip,
 amsmath,amssymb,
 reprint,%
floatfix]{revtex4-1}

\usepackage{graphicx}
\usepackage{dcolumn}
\usepackage{bm}

\usepackage[utf8]{inputenc}
\usepackage[T1]{fontenc}
\usepackage{mathptmx}
\usepackage{etoolbox}
\usepackage{hyperref}
\usepackage{xcolor}
\usepackage[caption=false]{subfig}

\makeatletter
\def\@email#1#2{%
 \endgroup
 \patchcmd{\titleblock@produce}
  {\frontmatter@RRAPformat}
  {\frontmatter@RRAPformat{\produce@RRAP{*#1\href{mailto:#2}{#2}}}\frontmatter@RRAPformat}
  {}{}
}%
\makeatother

\begin{document}

\preprint{AIP/123-QED}

\title[Many-body contributions to polymorphism and polyhexaticity in a water monolayer.]{Many-body contributions to polymorphism and polyhexaticity in a water monolayer.}

\author{Oriol Vilanova}
\affiliation{  F\'isica Estad\'istica i Interdisciplin\`aria--Departament de F\'isica de la Mat\`eria Condensada, Facutat de F\'isica, University of Barcelona, Mart\'i i Franqu\`es 1, Barcelona 08028, Spain}
\affiliation{Institute of Nanoscience and Nanotechnology (IN2UB), University of Barcelona, Mart\'i i Franqu\`es 1, Barcelona 08028, Spain}%

\author{Giancarlo Franzese}
\affiliation{  F\'isica Estad\'istica i Interdisciplin\`aria--Departament de F\'isica de la Mat\`eria Condensada, Facutat de F\'isica, University of Barcelona, Mart\'i i Franqu\`es 1, Barcelona 08028, Spain}
\affiliation{Institute of Nanoscience and Nanotechnology (IN2UB), University of Barcelona, Mart\'i i Franqu\`es 1, Barcelona 08028, Spain}%

\email{gfranzese@ub.edu}

\date{\today}

\pacs{}

\begin{abstract}
Nanoconfined water plays a crucial role in nanofluidics, biology, and cutting-edge technologies. The process of melting water monolayers and quasi-two-dimensional confined water involves, as an intermediate stage, the hexatic phase—a state that lies between solid and liquid and is characterized by quasi-long-range orientational order and short-range translational order. However, the influence of hydrogen bond (HB) cooperativity in this process has not been thoroughly investigated. This gap hampers our understanding of the phase behavior of confined water and limits the accuracy of our models. To address this, we extend the water model developed by Franzese and Stanley, which explicitly includes many-body interactions (MBIs) of HBs. We distinguish the contributions of three-body and five-body HB-MBIs. Our Monte Carlo calculations in the isobaric–isothermal ensemble produces a detailed pressure–temperature phase diagram, revealing polymorphism and polyhexaticity: low-density square ice and high-density triangular ice are separated from the liquid phase by distinct hexatic phases. Three-body interactions notably promote crystallization and can destabilize the low-density hexatic phase, while cooperative five-body interactions help restore it, thus modifying the thermodynamic landscape. These findings demonstrate that HB-MBIs are key in determining the phase behavior of confined water, influencing phenomena such as the non-monotonic specific heat, maximum density lines, and the accessibility of the liquid–liquid critical point. Beyond advancing theoretical understanding, these results have wide-ranging implications for nanofluidics, interfacial science, and applications in biology, food technology, and pharmaceutics, where controlling water under confinement is essential.

\end{abstract}

\maketitle

\section{Introduction}

Nanoconfined water plays a critical role in nanofluidics \cite{Nair2012}, biology \cite{Franzese2011, Bianco2012}, food science \cite{Aw2008}, and pharmaceutics \cite{Marti2017a}, and often serves as a benchmark for fundamental questions in physics \cite{Paolantoni2019}. Its behavior differs markedly from that of bulk water due to reduced dimensionality and surface interactions. Notably, two-dimensional (2D) confinement induces structural and dynamical anomalies that challenge existing understanding of water’s phase behavior. Recent research indicates that water confined between atomically flat surfaces can form layered ice structures, including square ice in monolayers and more complex arrangements in bilayers and trilayers \cite{Algara-Siller2015, Corsetti2016}.

The KTHNY theory \cite{Kosterlitz1972, Kosterlitz1973, Nelson1977, Halperin1978, Young1979} predicts that, in 2D, a solid exhibiting quasi-long-range translational order and long-range orientational order is separated from the liquid by two continuous phase transitions, with an intermediate hexatic phase characterized by quasi-long-range orientational order and the absence of translational order. The presence of the hexatic phase has been confirmed by large-scale simulations involving spherically symmetric interactions \cite{Chen1995}. However, its study remains technically challenging because it occurs within a very narrow parameter range \cite {Tsiok2018}, and the specifics of the interactions can influence how the solid and liquid phases are separated. For instance, although the KTHNY theory predicts two continuous transitions—solid-hexatic and hexatic-liquid—simulations of isotropic models, ranging from hard disks \cite{Bernard2011, Engel2013} and collapsing disks \cite{Ryzhov2017} to binary mixtures \cite{Russo2017}, as well as experiments with monolayers of colloidal hard spheres \cite{Thorneywork2017}, indicate a first-order transition between the hexatic and liquid phases. Regarding core-softened potentials, the hexatic-liquid transition can be discontinuous at low densities (LD) and continuous at high densities (HD) \cite{Zu2016}, or continuous at LD and absent at HD \cite{Dudalov2014, Dudalov2014a}. Additionally, computer simulations of soft colloidal Hertzian particles show multiple phase transitions consistent with both scenarios \cite{Fomin2018}. On the other hand, theory \cite{Chui1983} suggests that, for very short-range interactions \cite{Ryzhov2002}, crystal melting can occur via a first-order transition without the formation of a hexatic phase. Gribova et al. \cite{Gribova2011} demonstrated that strict confinement is necessary to stabilize hexatic phases, which tend to vanish in multilayer systems or under weak wall interactions. Russo and Wilding \cite{Russo2017} found that the hexatic phase in complex disk systems is highly fragile, disappearing at small impurity concentrations due to entropy-driven fractionation effects. These observations underscore the sensitivity of hexatic order to disorder and compositional perturbations.

Experimental evidence for hexatic phases has been observed in colloidal monolayers and smectic liquid crystals, where sixfold modulations in the structure factor reveal persistent orientational correlations \cite{Deutschlander2015}. Deutschländer et al. \cite{Deutschlander2013} verified the KTHNY scenario in colloidal monolayers with quenched disorder, demonstrating that the fluid–hexatic transition remains largely unchanged, while the hexatic–solid transition occurs at lower temperatures, expanding the hexatic stability range. These findings support the notion that orientational order can persist even in environments characterized by structural frustration.

In the case of water, hydrophobic nanoconfinement shifts the melting temperature and alters the phase diagram \cite{Zubeltzu2016a, Marti2017a}. Experiments involving water in narrow slit pores made of graphene-based membranes demonstrate the formation of mono- or multilayer square ice \cite{Algara-Siller2015}, consistent with simulation results \cite{Zangi2003, Zangi2003a}. Additional simulations with TIP5P water reveal polymorphic ice bilayers, with a first-order transition to honeycomb ice at low density and a continuous transition to rhombic ice at high density \cite{Han2010}. Further studies show that at high densities and elevated temperatures, water freezes into triangular ice via a continuous transition that includes an intermediate hexatic phase \cite{Zubeltzu2016a}. Although these findings, based on pairwise additive models, do not explicitly consider many-body interactions (MBI), calculations of the lowest-energy structures for empirical potentials incorporating polarization \cite{Hernandez-Rojas2010} indicate that clusters of up to 20 molecules tend to adopt fused-cube geometries rather than fused pentagonal arrangements, as predicted by non-polarizable potentials. Moreover, a nearby carbon-based material strongly stabilizes these cubic configurations. This supports the well-established understanding that MBIs influence water's properties \cite{Barnes1979}, although it does not help address the challenge of accurately incorporating them into atomistic models \cite{Cisneros2016}.

On the theoretical front, machine-learning (ML) potentials \cite{10.1063/5.0201241} derived from first-principles have been employed to generate comprehensive pressure–temperature phase diagrams for monolayer water \cite{Ravindra2024, Lin:2023aa}. These diagrams reveal a complex polymorphism, including a hexatic intermediate consistent with a two-step melting process, and a high-pressure regime characterized by increased proton mobility and partial ionic character \cite{Kapil2022}. Complementary large-scale path-integral molecular dynamics studies underscore the crucial role of nuclear-quantum effects and proton dynamics: for both monolayer and bilayer water, these studies identify numerous distinct 2D ice phases and a series of transitions from molecular to plastic, hexatic, and superionic-like states. These findings demonstrate that proton hopping and quantum motion significantly influence both the structure and transport properties in confined geometries \cite{Jiang2024}.

Experimental observations have revealed a new two-dimensional ice phase, known as square ice, that forms under ambient conditions when water is confined between graphene sheets \cite{Algara-Siller2015}. However, these findings have been debated \cite{Zhou2015}, suggesting that the observed lattice may be attributed to NaCl nanocrystals or contamination artifacts. This controversy underscores the need to develop theoretical models that explicitly incorporate many-body interactions (MBIs), which are crucial to water’s structural transitions \cite{Cisneros2016, Hernandez-Rojas2010}. In our study, we fill this gap by proposing a water model with adjustable MBIs and show that polyhexaticity and polymorphism in confined water arise from the intricate interplay between three-body and five-body HB contributions.

\section{THEORETICAL APPROACH}

\subsection{Water model with three and five-body interactions}

Simulations of water clusters using polarizable models demonstrate that including HB-MBIs within the first coordination shell (up to 5-body interactions) approximates the long-range MBIs to within 5\% \cite{Abella:2023ab}. This concept has recently been applied to develop a 3D water model that is reliable, efficient, scalable, and transferable (REST), and that accurately reproduces the thermodynamics and fluctuations of bulk water \cite{coronas2024phase, Coronas:2025aa, coronas2025algorithms}. The key advantage of this model is its simplicity, with very few parameters, two of which refer to the HBs: one ($J$) sets the HB covalent energy scale, and the other ($J_\sigma$) defines the HB-MBI energy scale. These parameters correspond to delocalization and polarization energies, respectively, as derived from the absolutely localized molecular orbital (ALMO) energy decomposition analysis (EDA) applied to density functional theory (DFT) calculations on water clusters \cite{KhaliullinKuhne2013}. This correspondence enables optimal parametrization, ensuring the model accurately captures HB properties fundamental to water thermodynamics \cite{Coronas:2025aa}.

To qualitatively analyze how various components of the MBIs influence water's melting behavior under two-dimensional confinement, we consider the Franzese-Stanley (FS) model for monolayers \cite{FSJPCM2002}. By focusing on HB properties and dynamics, the FS model enables detailed analysis of how various HB components influence water's anomalous behavior \cite{delosSantos2009, Stokely2010, delosSantos2012}.
Nevertheless, the FS model considers only a single energy scale, $J_\sigma$, for all many-body (or polarization) interactions. Here, we expand upon the original FS framework. Specifically, we introduce an additional parameter, $J_\theta$, that quantifies the energy scale of three-body interactions responsible solely for the relative positions of three nearest-neighboring (n.n.) molecules. This energy scale is distinct from $J_\sigma$, which is linked to any MBI up to five-body, and the two add up.

We consider a water monolayer confined between two parallel, hydrophobic plates separated by  $h \lesssim 1$~nm, with periodic boundary conditions applied laterally. The interaction between water molecules and the confining walls is modeled as purely repulsive. In its discretized \cite{FSJPCM2002} and continuous (off-lattice) formulations \cite{FranzesePRE2003}, the model effectively captures key thermodynamic and dynamic properties of water \cite{delosSantos2011}, and predicts polyamorphism \cite{Bianco2014}, aligning with other models \cite{Kesselring2012, Palmer2014} and experimental results \cite{Sellberg2014}. 

At any fixed temperature $T$ and pressure $P$, we partition the total occupied volume $V(T, P)$ between the two confining plates into a single plane of $N$ cells, each with height $h$ and volume $v_i \geq v_0$, where $v_0$ is the van der Waals excluded volume of a water molecule and $i\in[1, N]$. Following the original FS model, we assume that each $v_i$ contains a single water molecule and includes a homogeneous component $V_0/N$, where $V_0=V_0(T, P)$ is the volume in the absence of HBs, along with local inhomogeneities arising from the HBs the water molecule forms with its surroundings. In the following, we present further details on the definition of HBs and the relationship between their number $N_{\rm HB}$ and the total volume $V$.

The Hamiltonian of our model includes four energy terms:
\begin{equation}
\mathcal{H} \equiv \mathcal{H}_{\epsilon} + \mathcal{H}_{J} + \mathcal{H}_{5} + \mathcal{H}_{3}.
\label{H_FS}
\end{equation}
The first three terms are as in the continuous (off-lattice) formulation of the FS model \cite{FranzesePRE2003} recently reviewed in Ref.\cite{CoronasBook2022}. The fourth term is related to the three-body energy scale $J_\theta$ introduced here, as described below.

The first term, $\mathcal{H}_{\epsilon}$, features an isotropic 
Lennard-Jones interaction, which accounts for dispersive (van der Waals) attractive forces and short-range repulsive forces arising from the Pauli exclusion principle:
\begin{equation}
\mathcal{H}_{\epsilon} \equiv 4\epsilon \sum_{i < j} \left[ \left(\frac{r_0}{r_{ij}}\right)^{12} - \left(\frac{r_0}{r_{ij}}\right)^{6} \right],
\label{LJ}
\end{equation}
where $r_0 = 2.9$~Å is the van der Waals diameter \cite{Narten1967, SoperPRL2000}, 
and $4\epsilon = 5.8$~kJ/mol is the van der Waals attraction energy, 
consistent with the estimate of 5.5~kJ/mol based on isoelectronic 
molecules at optimal separation \cite{Henry2002} (Table \ref{Table1}).
The distance $r_{ij}$ is defined between the centers of two cells containing water molecules $i$ and $j$, respectively.
In the FS model, $r_{ij}$ does not depend on the HB's formation \cite{CoronasBook2022}, consistent with the experiments \cite{SoperPRL2000}. Hence, van der Waals interactions are unaffected by HBs and determine the homogeneous component $V_0(T, P)$ of the total volume. 
As we will discuss next, the formation of HBs introduces local inhomogeneities in the water molecules, affecting the volume each water molecule occupies, but not the distance to its first hydration shell. 

The sum in Eq.(\ref{LJ}) runs over all pairs $(i,j)$ within a cutoff distance $r_{\rm cut} = 2.5 r_0$, introduced to reduce computational cost. The interaction is offset by a constant such that $\mathcal{H}_{\epsilon}$ vanishes for $r_{ij} \ge r_{\rm cut}$. Since the system is simulated in the isobaric-isothermal ensemble (constant $N$, $P$, $T$), $V_0$ and $r_{ij}$ are continuous variables. For this reason, the FS model that adopts the isotropic term in Eq.(\ref{LJ}) is called {\it continuous}.

The second term in Eq.(\ref{H_FS}), $\mathcal{H}_J \equiv -J N_{\rm HB}$, represents the directional two-body HB interaction \cite{Isaacs2000, Pendas2006}, where $N_{\rm HB}$ is the total number of HBs between water molecules. We set $J/4\epsilon=0.5$, i.e., $J=11.6$ kJ/mol, as estimated in \cite{Stokely2010} assuming the optimal HB energy $\approx 23.3$ kJ/mol \cite{Suresh2000} (Table \ref{Table1}). 

\begin{table}
\begin{tabular}{ |c|c|c| } 
 \hline
 Parameter & Internal units & Real units\\
 \hline
 $4\epsilon$ & 1 & 23.20 kJ/mol \\ 
 \hline
 $J$ & 0.5 & 11.60 kJ/mol\\ 
 \hline
 $J_{\sigma}$ & 0.1 & 2.32 kJ/mol\\
 \hline
 $J_{\theta}$ & 0.08 -- 0.10 -- 0.12 & 1.86 -- 2.32 -- 2.78 kJ/mol\\
 \hline
  $r_{0}$ & 1 & 2.9 Å\\
  \hline
 $v_{\rm{HB}}$ & 0.5$v_0$ & 6.4 Å$^3$ \\
 \hline
 $q$ & 6 & 6\\ 
 \hline
\end{tabular}
\caption{Model parameters. Values are set as in previous works for the FS model \cite{Bianco2014}, except for $J_{\sigma}$, which is twice as large as that used before, e.g., in Ref. \cite{Bianco2014}. As $J_{\sigma}$ increases, the model reaches equilibrium at low temperatures without necessitating lengthy simulations, and the qualitative outcomes stay consistent. However, the precise locations of phase boundaries for $T<J_{\sigma}/k_B$ depend on the specific value of $J_{\sigma}$, as explained in the text. The parameters are given in internal units ($4\epsilon$) and in real units, using the conversion method presented in Ref. \cite{Bianco2014}.
}
\label{Table1}
\end{table}

We adopt the standard HB definition based on the distance between molecular centers and the angle between the OH group and the oxygen atom in two bonded molecules. A HB reaches its minimum energy when the hydrogen atom lies along the O--O axis or deviates by less than $30^\circ$. Consequently, only one-sixth of the possible in-plane orientations correspond to a bonded state, while the remaining five-sixths do not. To accurately account for the entropy change during HB formation, we introduce a six-state bonding variable $\sigma_{ij}$ for each of the four potential HBs a water molecule $i$ can form with neighboring molecules $j$. A bond is established only when two facing molecules share the same bonding state. As a result, the total number of HBs is $N_{\rm{HB}} \equiv \sum_{\langle i,j\rangle}\delta_{\sigma_{ij},\sigma_{ji}}$, where $\delta_{\alpha,\beta}\equiv 1$ if $\alpha=\beta$, $0$ otherwise, and the sum runs over n.n., with each water molecule capable of forming up to four HBs. 

HB formation yields an open, locally tetrahedral structure that increases the volume per molecule \cite{SoperPRL2000, Soper2008}. To account for this effect, we assume a local volume increase of $v_{\rm HB} = 0.5 v_0$ per HB formed, consistent with the observed density difference between low-density (tetrahedral) ice Ih and the high-density ices VI and VIII in bulk water. Consequently, the total system volume is given by 
\begin{equation}
V \equiv V_0 + v_{\rm HB} N_{\rm HB}
\label{Vtot}
\end{equation}
where $V_0$ and $N_{\rm HB}$ denote the volume without HBs and the total number of HBs, respectively, as defined above.

The third term in Eq.(\ref{H_FS}), $\mathcal{H}_{5}$, accounts for cooperative MBIs among HBs formed by a single molecule with up to four other molecules in its first hydration shell. It is defined as
\begin{equation}
\mathcal{H}_{5} \equiv -J_{\sigma} \sum_{i} \sum_{(k,\ell)_i} \delta_{\sigma_{ik}, \sigma_{i\ell}}
\label{H5}
\end{equation}
where $J_{\sigma}$ is the energy gain associated with enhanced local orientational order, and $(k,\ell)_i$ indicates each of the six distinct pairs among the four bonding indices $\sigma_{ij}$ of molecule $i$. This term promotes local orientational coherence by favoring configurations where multiple HBs originating from the same molecule share the same bonding state.

The fourth term in Eq.(\ref{H_FS}), $\mathcal{H}_{3}$, is absent from the FS model and is introduced here for the first time. It accounts for three-body HB interactions and introduces angular dependence among triplets of bonded molecules $i$, $k$, and $\ell$. To define the angle of a triplet, we use the center-of-mass coordinates of each molecule $i$, $k$, and $\ell$.
This term favors square-like arrangements over close-packed configurations and is defined as
\begin{equation}
\mathcal{H}_{3} \equiv J_{\theta} \sum_{i} \sum_{\langle \langle k,\ell \rangle \rangle_i} \delta_{\sigma_{ik}, \sigma_{ki}} \, \delta_{\sigma_{i\ell}, \sigma_{\ell i}} \, \Delta(\theta^{i}_{k\ell}),
\label{H3}
\end{equation}
where $J_{\theta}$ is the coupling constant for three-body interaction, and $\langle \langle k,\ell \rangle \rangle_i$ denotes all neighboring pairs of molecules $k$ and $\ell$ bonded to a central molecule $i$, with the constraint that $k$ and $\ell$ are second n.n. to each other, and $\theta^{i}_{k\ell}$ is the angle formed by $i$, $k$, and $\ell$.
We choose
\begin{equation}
\Delta(\theta) \equiv \frac{1}{2} \left[ 1 - \cos(4\theta) \right],
\label{delta_theta}
\end{equation}
which is a smooth angular function with minima at $\theta = \{ 0^{\circ}, 90^{\circ}, 180^{\circ} \}$ for $\theta \leq 180^{\circ}$.
This term penalizes deviations from square symmetry and energetically disfavors triangular configurations (with $\theta \approx 60^{\circ}$). A detailed analysis of this term and its implications is presented in the following section.

\subsection{Computational Methods}

We simulate systems in the isobaric-isothermal ensemble (NPT), i.e., we keep the pressure $P$, temperature $T$, and number of molecules $N$ constant during a simulation, while the total volume $V$ is a continuous variable, as are the coordinates of the molecules $r_{ij}$. Therefore, the volume per molecule $v$ and the number density $\rho\equiv 1/v$ are functions of the control parameters $N$, $P$, and $T$. In this work, we present the results of simulations performed with the set of model parameters described in Table \ref{Table1}. Further details of the simulation protocol are provided in the Supplementary Information (SI, \href{./supplementary.pdf}{I}).

To characterize the structure of the liquid, we calculate the 
radial distribution function (RDF). 
The RDF is defined as 
\begin{equation}
	g(r) \equiv \left\langle \frac{1}{N\rho} \sum_{i \neq j} \delta(r - r_{ij}) \right\rangle,
\label{gr}
\end{equation}
 where the symbol $\langle \dot \rangle$ represents an average  over independent configurations.
We identify the system's phases through a comprehensive structural analysis that involves calculating multiple orientational and translational order parameters. Detailed methodological information is provided in the SI (\href{./supplementary.pdf}{II}).

Furthermore, to examine the thermodynamics of the system, we compute the heat capacity at constant pressure, $C_P$, as the statistical fluctuations of the enthalpy $H$:
\begin{equation}
C_P =  \frac{1}{k_B T^2}\left(\langle H^2 \rangle - \langle H \rangle^2\right),
\label{Cp}
\end{equation} 
where $H \equiv U + PV$,  
$U\equiv \langle \mathcal{H} \rangle$ in Eq.(\ref{H_FS}), 
and $V$ is the total volume in Eq.(\ref{Vtot}).

\section{RESULTS}

For the model with parameters set as in Table \ref{Table1}, we show that the confined water monolayer exhibits polymorphism, with two distinct ice forms at low temperatures. 
We also observe polyhexaticity: each solid phase is separated from the liquid by a corresponding hexatic phase (Fig. \ref{fig:PhaseDiagram}).

\subsection{Phase Diagram}

\begin{figure}
\includegraphics[angle=0, width=\linewidth]{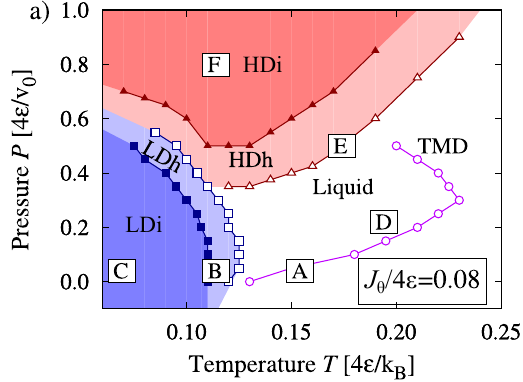}\vspace{1em}
\includegraphics[angle=0, width=\linewidth]{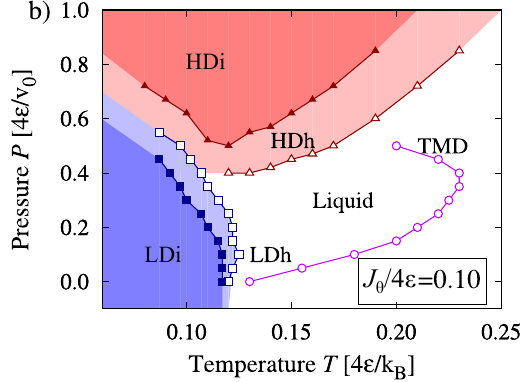}\vspace{1em}
\includegraphics[angle=0, width=\linewidth]{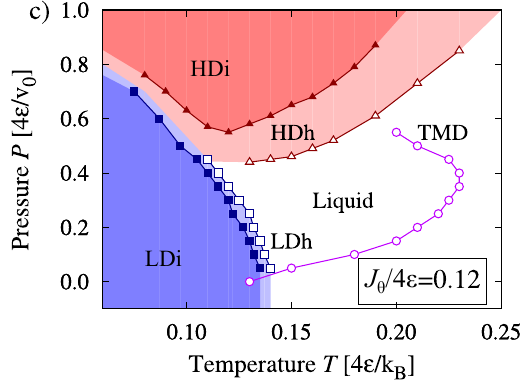}
\caption{{\bf Polymorphism and polyhexaticism in a water monolayer.}
$P-T$ phase diagram with parameters $J/4\epsilon=0.5$,  $J_{\sigma} /4\epsilon= 0.1$ and for: 
{\bf (a)} $J_{\theta}/4\epsilon= 0.08$, 
{\bf (b)} $J_{\theta}/4\epsilon= 0.10$, and 
{\bf (c)} $J_{\theta}/4\epsilon= 0.12$. 
Other parameters are detailed in Table \ref{Table1}. We identify the liquid (white region), LD hexatic (LDh, pale blue), HD hexatic (HDh, pale red), LD ice (LDi, blue), and HD ice (HDi, red) phases, as well as the line of temperatures of maximum density (TMD, line with open circles) in the liquid phase. The gas phase at higher temperatures is not shown. The boundaries of the hexatic phases (marked by lines with open squares and triangles) are determined through structural analysis, as detailed in SI (\href{./supplementary.pdf}{II.C}). Letters in {\bf (a)} denote the state points considered in text.} 
\label{fig:PhaseDiagram}
\end{figure}

To investigate the impact of varying the three-body HB interaction strength, we analyze the system's behavior for three different coupling constants: $J_\theta/4 \epsilon = 0.08$, $0.10$, and $0.12$. For each value, we generate the pressure–temperature ($P$–$T$) phase diagram (Fig.\ref{fig:PhaseDiagram}) and identify the associated thermodynamic phases: low-density ice (LDi), low-density hexatic (LDh), high-density ice (HDi), high-density hexatic (HDh), and liquid. The diagrams also include the temperature of maximum density (TMD) line and the boundaries of the hexatic regions, determined through structural analysis. This comparative analysis enables us to evaluate how increasing $J_\theta$ affects the stability and extent of each phase, especially the hexatic phases and the transitions between solid and liquid states.

First, we fix $J_{\theta}/4 \epsilon=0.08$ and observe that the water monolayer remains liquid within the considered pressure range $P$ at sufficiently high temperatures $T$ (white region in Fig.\ref{fig:PhaseDiagram}a). Subsequently, we increase the three-body parameter $J_{\theta}/4 \epsilon$ to 0.10 and 0.12 (Fig.\ref{fig:PhaseDiagram}b and c). We find that the LD solid and hexatic phases expand to higher $T$ and $P$ as $J_{\theta}$ increases, with a more pronounced effect on the solid, resulting in a narrower LDh. Conversely, the impact on HD 
phases and the overall topology of the phase diagram remains minor within the explored $J_{\theta}$ range. Hence, we observe both polymorphism, evidenced by the LDi and HDi phases, and polyhexaticism, indicated by LDh and HDh phases. The LD phases are reentrant in temperature $T$, whereas the HD phases are reentrant in pressure $P$. Near the high-$T$ limit of the LD phases, increasing pressure drives the system from LD to HD phases by crossing the liquid phase. Conversely, near $Pv_0/4\epsilon=0.6$, increasing $T$ induces a sequence of transitions: from LDi to LDh, then to HDh, followed by HDi, back to HDh, and ultimately to the liquid phase.

\underline{\emph{TMD line.}} Within the liquid region of each phase diagram (white areas in Fig.\ref{fig:PhaseDiagram}), we identify a line of temperatures of maximum density (TMD) at constant $P$, corresponding to anomalous density behavior of water under confinement, similar to that observed in bulk. For all three values of $J_\theta$, the TMD line shows a negative slope in the $P$–$T$ plane, consistent with experimental findings for bulk water at pressures above $-137$~MPa \cite{Caldwell1978, Henderson1987, Holten2017}. At lower pressures, the TMD line bends and changes slope, as observed in simulations of atomistic models of bulk water \cite{llcp, Holten2017}. As $J_\theta$ increases, the TMD turning-point shifts slightly to higher $P$, yet the line retains its overall shape and slope, suggesting that the three-body interaction has limited influence over the temperature range of the TMD line.

\underline{\emph{Low-density phases.}} 
At low pressures and temperatures, the system displays two distinct low-density phases: a square ice phase (LDi) and a low-density hexatic phase (LDh). When $J_\theta/4\epsilon = 0.08$ (Fig.\ref{fig:PhaseDiagram}a), the LDh phase appears between the liquid and LDi phases, characterized by quasi-long-range orientational order and short-range translational order. Further details on the phase transition lines are provided in the SI. As $J_\theta$ increases (Fig.\ref{fig:PhaseDiagram}b, c), the LDh region narrows and is eventually replaced by the LDi phase, indicating that stronger three-body interactions promote the formation of square ice over intermediate hexatic ordering. This dependence underscores the importance of angular constraints in stabilizing low-density crystalline structures. The LDi phase also expands in both temperature and pressure with increasing $J_\theta$, reflecting an enhanced energetic preference for square arrangements.

\begin{figure}
 \includegraphics[width=\linewidth]{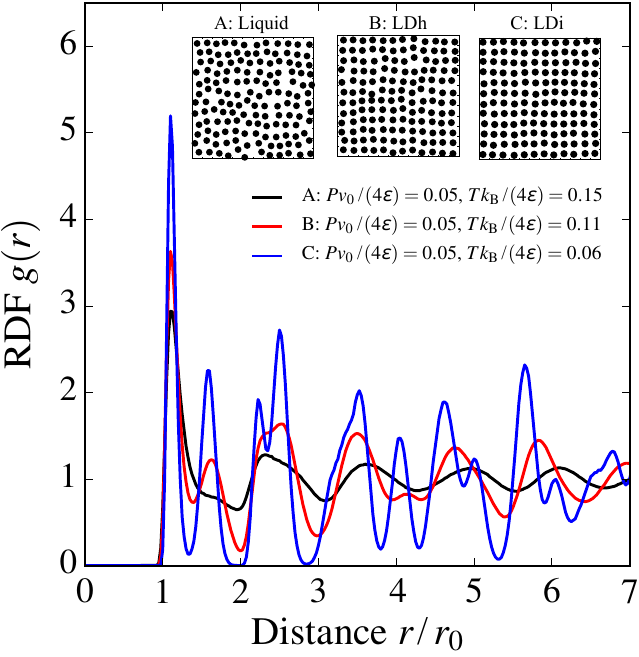}
 \includegraphics[width=\linewidth]{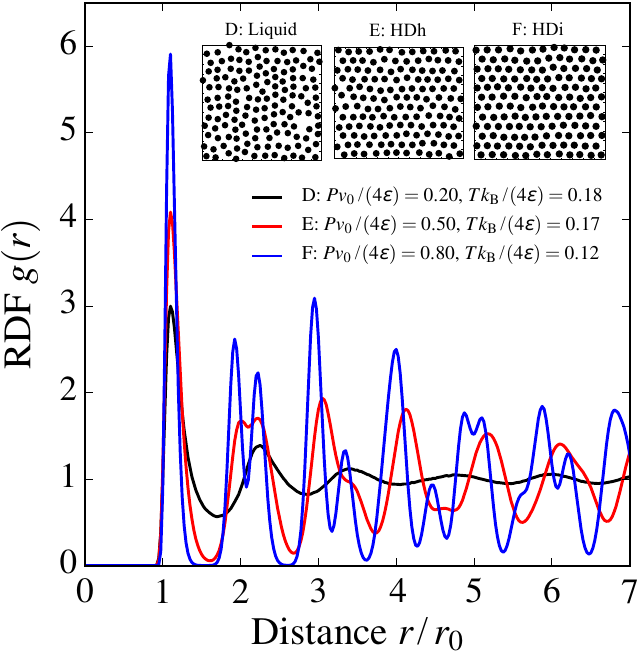}
 \caption{{\bf Structural changes in a water monolayer at different state-points.}
Radial distribution function $g(r)$  for  the model with $J_{\theta}/4\epsilon =0.08$  evaluated  at:
{\bf (Top)}   $Pv_0/4\epsilon=0.05$ and 
 (A) $Tk_B/4\epsilon=0.15$ in the liquid phase, 
 (B) $Tk_B/4\epsilon=0.11$ in the LDh phase and 
 (C) $Tk_B/4\epsilon=0.06$ in the LDi phase;  
{\bf (Bottom)} 
at
(D)  $Pv_0/4\epsilon=0.2$ and $Tk_B/4\epsilon=0.18$ in the liquid phase, 
(E)   $Pv_0/4\epsilon=0.5$ and $Tk_B/4\epsilon=0.17$ in the HDh phase and 
(F) $Pv_0/4\epsilon=0.8$ and $Tk_B/4\epsilon=0.12$ in the HDi phase.
Insets: Configurations of the molecules' center of masses corresponding to the state points {\bf (Top)} A, B, C, and {\bf (Bottom)} D, E, F. The letters refer to the state-points marked in Fig.\ref{fig:PhaseDiagram}a.}
\label{fig:RDF1}
\end{figure}

\underline{\emph{High-density phases.}} At high pressures and low temperatures, the system exhibits two high-density phases: a triangular ice phase (HDi) and a high-density hexatic phase (HDh). The HDh phase is distinguished by sixfold orientational order and short-range translational order, aligning with predictions from the KTHNY theory (see SI \href{./supplementary.pdf}{II.C} for details). Unlike the LDh phase, the HDh region remains stable across all values of $J_\theta$, suggesting that three-body interactions minimally influence the stability of hexatic order at high density. The HDi phase, characterized by triangular symmetry and reduced hydrogen bonding, expands slightly with increasing $J_\theta$, though its boundaries stay largely constant. Notably, both HD phases show reentrant behavior in pressure: the HDh phase reappears at higher temperatures after the HDi phase melts.

\subsection{Structural Analysis}

To characterize the structural evolution across the phase diagram, we analyze the radial distribution function (RDF) $g(r)$ for representative state points, focusing on the case of the three-body interaction parameter $J_{\theta}/4\epsilon=0.08$ (Fig.\ref{fig:RDF1}). 
At low pressure ($Pv_0/4\epsilon=0.05$), the RDF in the liquid phase ($Tk_B/4\epsilon=0.15$) exhibits a prominent peak at $r/r_0=1$, with smooth oscillations decaying toward unity, indicating no long-range order. When the temperature decreases to $Tk_B/4\epsilon=0.11$, a secondary peak appears around $r/r_0 \simeq 1.7$, along with a shoulder near $r/r_0 \simeq 2.4$, reflecting quasi-long-range orientational order and short-range translational order characteristic of the LDh phase. Corresponding molecular configurations reveal fourfold symmetry. Further cooling to $Tk_B/4\epsilon=0.06$ produces sharp peaks and minima approaching zero, characteristic of square crystalline order in the LDi phase.

At higher pressure ($Pv_0/4\epsilon = 0.2$), the RDF in the liquid phase ($Tk_B/4\epsilon = 0.18$) becomes less structured, with decreased peak intensity and damped oscillations, reflecting increased thermal disorder. Increasing the pressure to $Pv_0/4\epsilon = 0.5$ and lowering the temperature to $Tk_B/4\epsilon = 0.17$ results in prominent peaks and sustained oscillations in the RDF, indicating the onset of the HDh phase with sixfold orientational symmetry. The RDF reveals splitting of peaks and shoulder development, such as for thepeaks at $r/r_0 \simeq 2.1$ and the shoulder at $r/r_0 \simeq 3.2$, consistent with the liquid–hexatic transition \cite{Truskett1998}. At even higher pressure ($Pv_0/4\epsilon = 0.8$) and reduced temperature ($Tk_B/4\epsilon = 0.12$), the RDF exhibits a highly ordered triangular arrangement with multiple sharp peaks and deep minima, confirming the formation of the HDi phase. These RDF features enable clear differentiation between liquid, hexatic, and solid phases, aligning with structural classifications based on symmetry and correlation length. Complemented by visual inspection of molecular configurations and symmetry analysis, the data confirm the presence of square and triangular lattices in the LDi and HDi phases, respectively.

Near the TMD line at $Pv_0/4\epsilon=0.05$ and $Tk_B/4\epsilon=0.15$ (line A in Fig.\ref{fig:RDF1}), the RDF exhibits a prominent peak at $r/r_0 = 1$. For larger r values, the RDF shows several oscillations approaching 1, consistent with a liquid state characterized by no long-range order (inset A in Fig.\ref{fig:RDF1}). When $Tk_B/4\epsilon$ is decreased to 0.11 (line B in Fig.\ref{fig:RDF1}), a new peak emerges at $r/r_0 \simeq 1.7$, along with a shoulder at $r/r_0 \simeq 2.4$. The typical configuration at this state point (inset B in Fig.\ref{fig:RDF1}) indicates quasi-long-range orientational order and short-range translational order, characteristic of the hexatic phase as discussed in SI. This phase exhibits fourfold symmetry and is here referred to, for simplicity, as the low-density hexatic (LDh), although a more precise designation is discussed in the Methods. At an even lower temperature, $Tk_B/4\epsilon = 0.06$, the RDF displays several new peaks and minima approaching zero (line C in Fig.\ref{fig:RDF1}), closely resembling the square crystal case (SI). The typical configuration at this point (inset C in Fig.\ref{fig:RDF1}) corresponds to low-density ice (LDi).

Next, we compute the RDF at the state point $Pv_0/4\epsilon = 0.2$ and $Tk_B/4\epsilon = 0.18$ (line D in Fig.\ref{fig:RDF1}). The main peak appears at $r/r_0 = 1$, with small oscillations around 1 for larger $r$, indicative of a liquid-like structure (inset D in Fig.\ref{fig:RDF1}). Notably, the RDF at point D exhibits less structure compared to that at A, consistent with the higher temperature of point D relative to A \cite{Hoang2016}. Increasing the pressure to $Pv_0/4\epsilon = 0.5$ and lowering the temperature to $Tk_B/4\epsilon = 0.17$ (line E in Fig.\ref{fig:RDF1}) results in multiple peaks and pronounced oscillations around 1. Our detailed analysis of spatial correlations (SI) indicates that point E corresponds to a high-density hexatic (HDh) phase, characterized by sixfold orientational order (inset E in Fig. \ref{fig:RDF1}). Further increasing the pressure and decreasing the temperature (line F in Fig.\ref{fig:RDF1}) yields a highly structured RDF with many peaks and minima approaching zero, consistent with the symmetry of a triangular lattice \cite{Mathews2017}. Structural analysis (SI) confirms that the system forms a triangular crystal (inset F in Fig. \ref {fig:RDF1}), which we denote as high-density ice (HDi).

\subsection{Specific Heat}

\begin{figure}
  \includegraphics[width=\linewidth]{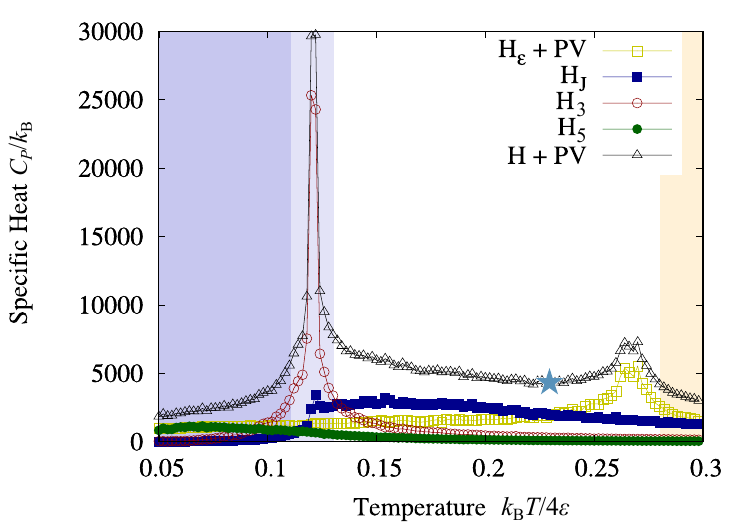}
\caption{{\bf Specific heat decomposition of a water monolayer at low pressure}.
For $J_{\theta}/4\epsilon=0.08$ at $Pv_0/4\epsilon =0.05$, we calculate the specific heat $C_P$ (dashed line with gray triangles) 
from enthalpy ($\mathcal{H} +PV$) fluctuations and decompose it into its components:
i)  the van der Waals component $\mathcal{H}_{\epsilon} +PV$ (yellow open squares);
ii)  the two-body HB component $\mathcal{H}_{J}$ (blue squares);
iii)  the three-body component $\mathcal{H}_{3}$ (green circles);
iv) the first coordination shell (up to) five-body component $\mathcal{H}_{5}$ (red open circles).
The lines are guides for the eye. The star marks the minimum in $C_P$. 
The shaded regions mark the LD ice (blue), LD hexatic (pale blue), 
the liquid (white), and the gas phase (yellow).}
\label{fig:spHeat}
\end{figure}

To explore the thermodynamic behavior of the system, we examine the isobaric specific heat, $C_P$, as a function of temperature across various pressures and values of $J_\theta$. For $J_\theta/4\epsilon = 0.08$, at low pressure $Pv_0/4\epsilon  = 0.05$ (Fig. \ref{fig:spHeat}), $C_P$ displays a non-monotonic pattern with two distinct maxima. The two primary maxima occur at distinct temperatures: a broad peak in the liquid phase at $Tk_B/4\epsilon \simeq 0.16$, and a sharper peak within the LDh phase at $Tk_B/4\epsilon \simeq 0.12$. These peaks merge into a single asymmetric maximum in $C_P$, characterized by a steep low-temperature tail and a broader high-temperature shoulder, consistent with experimental observations of confined water \cite{Mallamace2014}. The decomposition of $C_P$ into its components reveals that the broad maximum is linked to the two-body hydrogen bond energy, $\mathcal{H}_J$, while the sharp peak results from fluctuations in the many-body component $\mathcal{H}_{5}$, which has an amplitude over six times greater than that of $\mathcal{H}_J$. Additionally, a weaker contribution from the three-body term, $\mathcal{H}_{3}$, appears at lower temperatures, with a broad maximum around $Tk_B/4\epsilon  \simeq 0.06$.

\begin{figure}
\includegraphics[width=\linewidth]{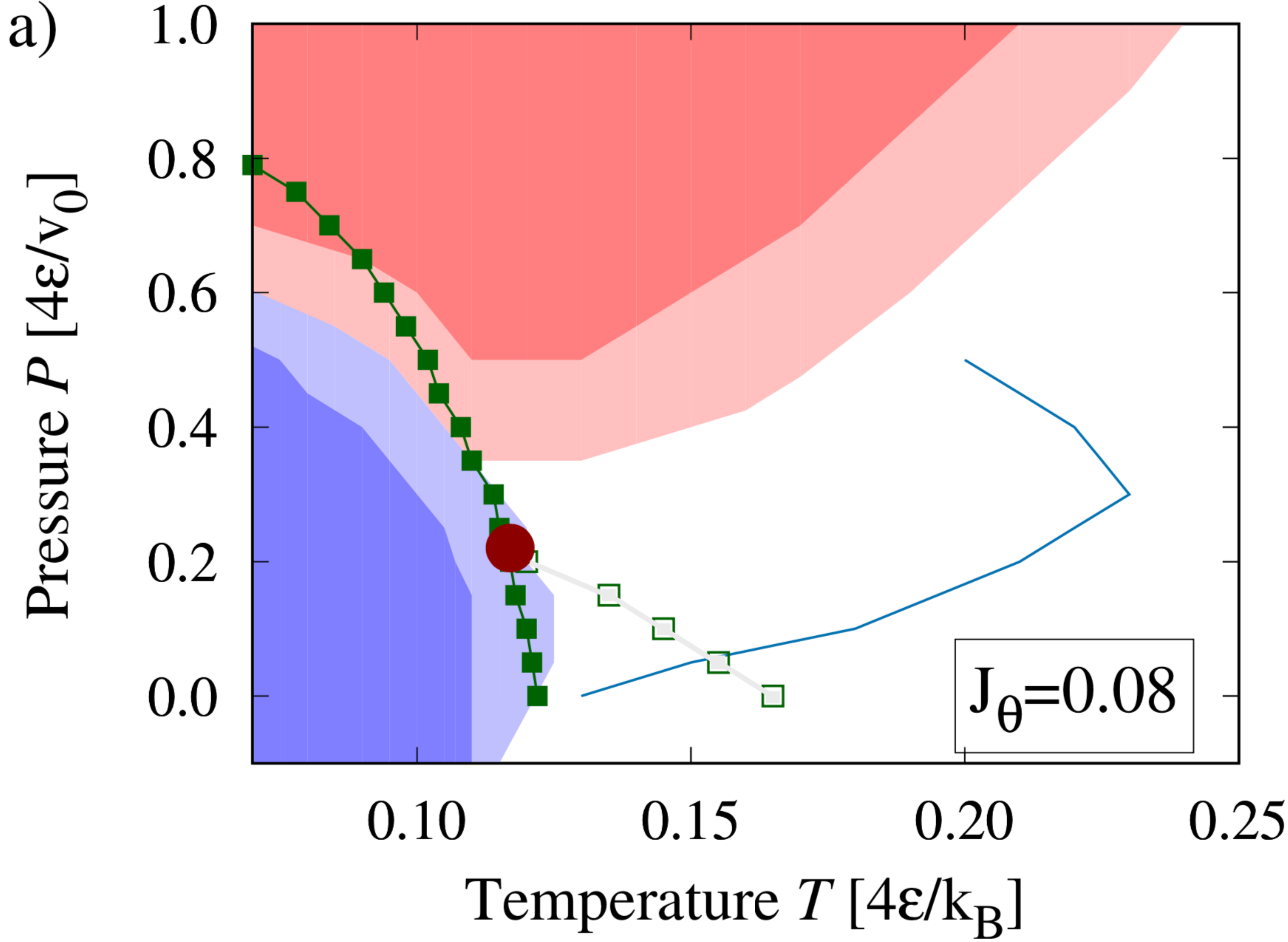}\vspace{1em}
\includegraphics[width=\linewidth]{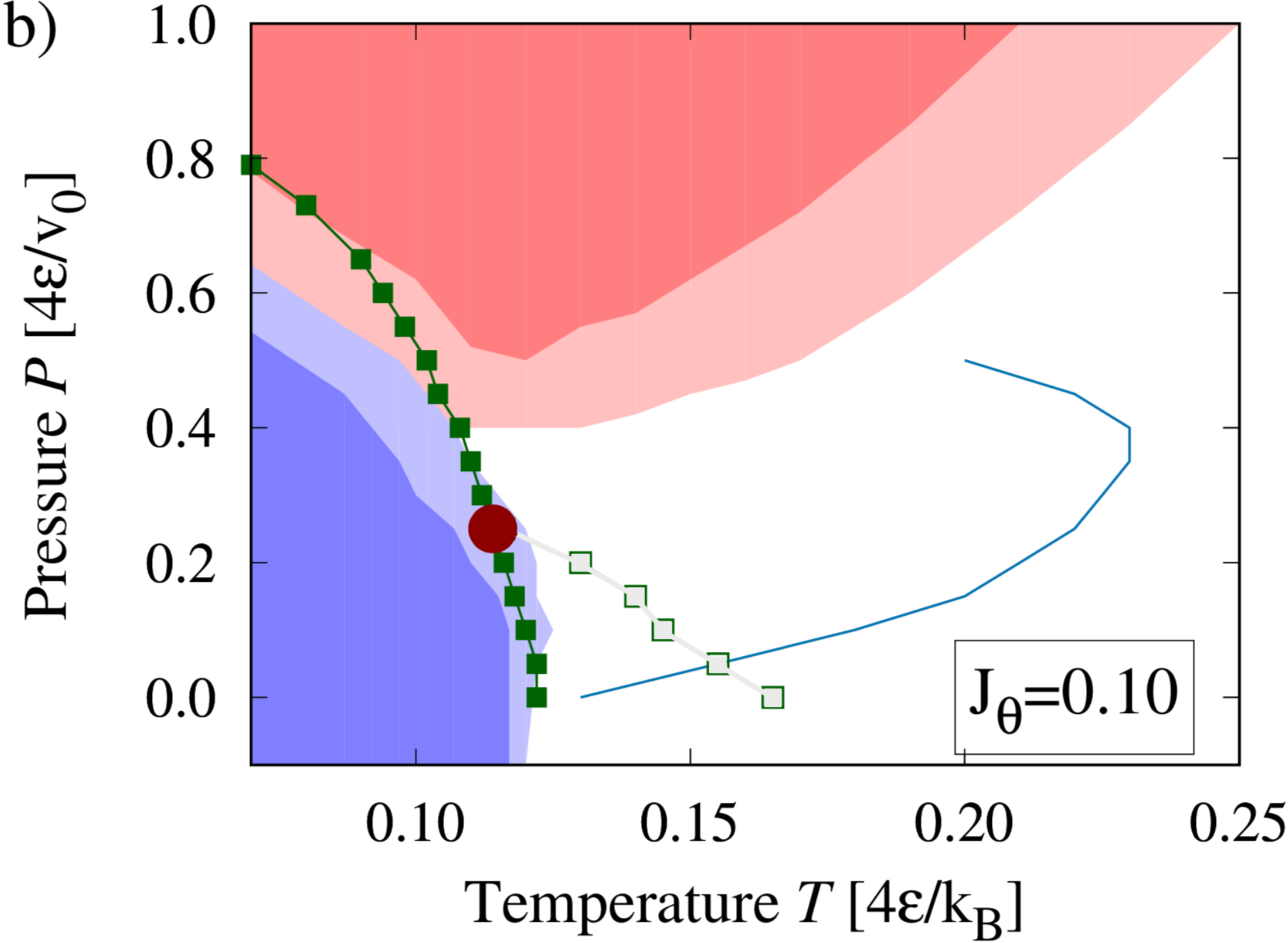}\vspace{1em}
\includegraphics[width=\linewidth]{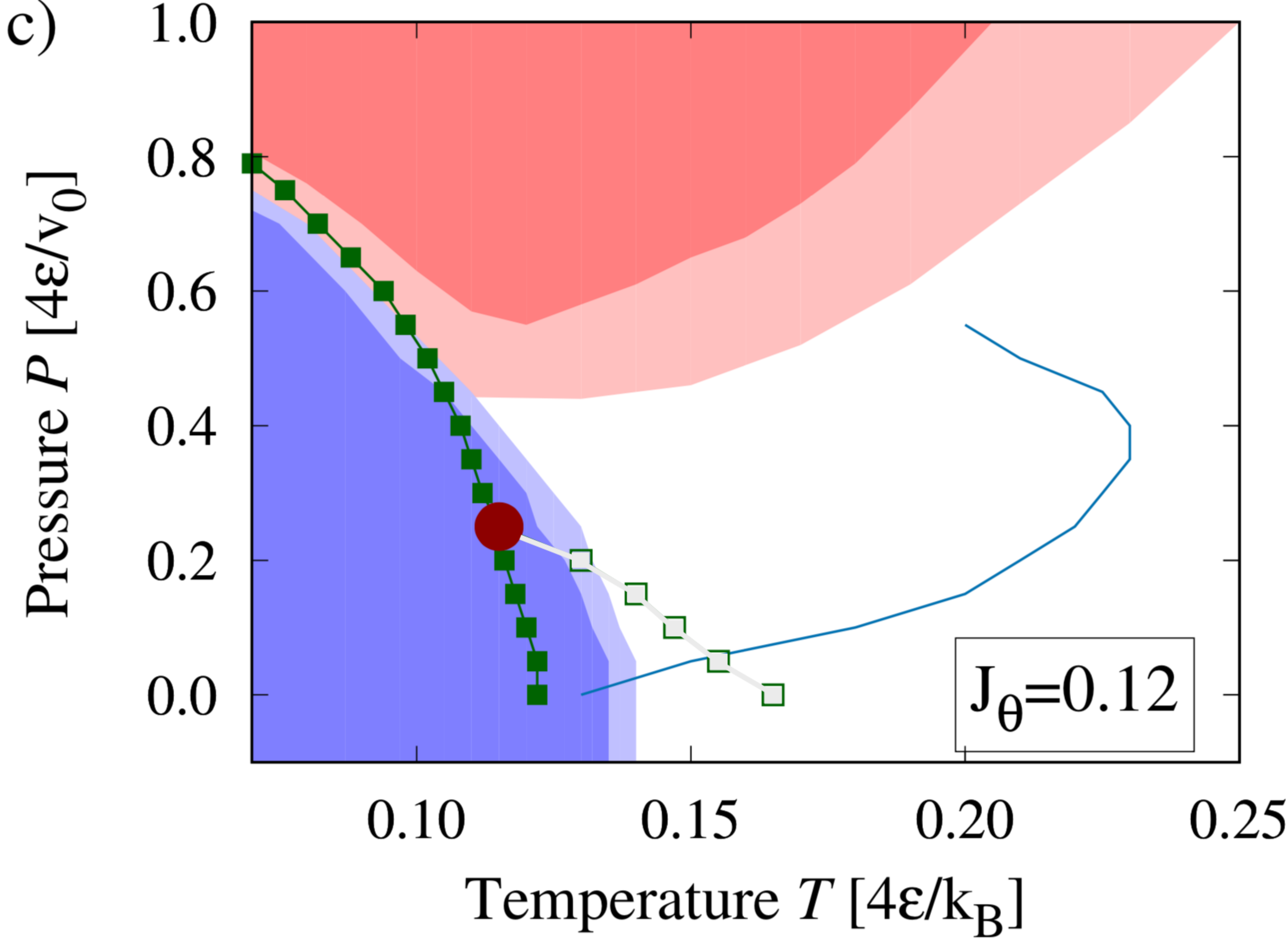}
\caption{{\bf  Locus of maxima of  specific heat in the $P$-$T$ phase diagram.} 
The line of $C_P$ maxima (green with full squares) 
and the maximum in the two-body HB component of $C_P$(green open squares with gray line) converges at $Pv_0/4\epsilon \simeq 0.25$ and $Tk_B/4\epsilon \simeq 0.115$  (purple circle), regardless of $J_{\theta}$. This point resembles a critical point at the end of a line of first-order phase transitions. The first-order phase-transition line coincides with the $C_P$-maximum line at high pressure. The TMD (blue) line is shown as a reference. Model parameters are as in Fig.\ref{fig:PhaseDiagram}: $J_{\theta}4\epsilon=0.08$ in (a), $0.10$ in (b), and $0.12$ in (c). The lines serve as guides to the eye.} 
\label{fig:SpecHeat}
\end{figure}

As $J_\theta$ increases (Fig. \ref{fig:SpecHeat}), the LDh region becomes more restricted, and the sharp peak shifts toward lower temperatures, while the broad maximum remains relatively stable. The line of $C_P$ maxima traces a path across the phase diagram, intersecting the LD and HD phases but not clearly entering the liquid region, converging toward the maxima of $\mathcal{H}_J$ fluctuations.
For larger $J_\theta$, the LD solid and hexatic phases encompass an increasing portion of the $C_P$ maxima locus, further highlighting the influence of many-body interactions on the thermodynamics of confined water. 

\subsection{Liquid–Liquid Critical Point and Many-Body Effects}

The convergence of the locus of maxima of $C_P$ with the maxima of $\mathcal{H}_J$ fluctuations occurs at high pressure, around $Pv_0/4\epsilon \simeq 0.25$ and $Tk_B/4\epsilon \simeq 0.115$ (Fig. \ref{fig:SpecHeat}). It remains consistent regardless of $J_\theta$ and resembles the merging of broad and sharp $C_P$ peaks observed in the FS model at the liquid–liquid critical point (LLCP) \cite{Bianco2014}, suggesting that a similar critical phenomenon may be present in our extended framework.

Previous studies have shown that various models of supercooled water predict a liquid–liquid phase transition (LLPT) culminating in an LLCP, driven by the competition between low- and high-density liquid structures. For bulk water, the LLCP falls within the 3D Ising universality class, while for confined water monolayers, it belongs to the 2D Ising universality class \cite{Bianco2014}. A crossover towards the 3D Ising class occurs in systems with lateral sizes comparable to the confining width \cite{Bianco2014}. 

In the current model, we observe thermodynamic signatures similar to those expected near the LLCP. However, including explicit three- and five-body HB interactions shifts the balance among competing structural motifs. For instance, a strong three-body term suppresses the LD hexatic phase and promotes crystallization, potentially hindering the LLCP. Conversely, the five-body interaction stabilizes orientational fluctuations and amplifies the LD–HD crossover. 

These findings suggest that the detectability of the LLCP is highly sensitive to the relative strengths of the many-body components and may be preempted by solidification or broadened into a crossover, depending on the parameter settings. In particular, the preempted scenario has been examined within the framework of the mW water model \cite{Moore2009}, in which water cooperativity is captured exclusively by a three-body term that promotes crystallization \cite{Chandler2011}, consistent with our findings. 

Our results highlight the importance of many-body interactions in determining the phase behavior of confined water. Although the LLCP is thermodynamically possible, it can be challenging to observe experimentally or even in simulations. In particular, models require careful balancing of many-body interactions and precise parameter tuning.

\section{Conclusions}

By incorporating explicit three- and five-body HB terms into the FS model, we show that many-body interactions (MBIs) play a fundamental role in the melting of two-dimensional water.
Our comprehensive study reveals a complex phase diagram exhibiting polymorphism, polyhexaticity, and thermodynamic anomalies. Overall, our findings underscore the importance of explicitly modeling many-body hydrogen-bond (HB) interactions to accurately capture the complex behavior of confined water. The interplay of directional bonding, angular constraints, and cooperative effects governs phase stability and transitions, providing valuable insights for developing transferable water models and informing future experimental and theoretical research. 

We perform MC calculations of the radial distribution function, finding distinct local arrangements and symmetry transitions, leading to a scenario in which both low-density (LD) ice and high-density (HD) ice are separated from the liquid by hexatic phases (LDh and HDh, respectively) characterized by the absence of translational order and quasi-long-range orientational order with four-fold and six-fold symmetry, respectively.

We observe that the LD ice and LDh phase are highly responsive to the relative strengths of the MBIs. For our chosen parameters, a 50\% relative increase in the three-body interaction energy promotes crystallization at low pressure and reduces the LDh phase between ice and liquid. In contrast, the HDh phase remains stable across all conditions, indicating that its order is less affected by angular constraints.

On the other hand, including MBIs up to five is sufficient to balance the effects of the three-body term. For example, a five-body interaction that is 25\% stronger than the three-body term is enough to reduce the temperature and pressure range for LD ice and to extend the LDh phase over a broader temperature range. 

Additionally, the thermodynamic analysis shows non-monotonic behavior in the specific heat $C_P$, with multiple peaks associated with different enthalpy components. The calculated specific heat closely resembles that observed in experiments with confined water \cite{Mallamace2014}, but only partially in bulk water \cite{Pathak2021}. This highlights that the impact of MBI on water can vary significantly with the boundary conditions. 

The convergence of the $C_P$ peak and the maximum in the two-body HB component of $C_P$ at a specific point in the phase diagram suggests a liquid–liquid critical point (LLCP), aligning with previous studies of the FS model \cite{Bianco2014}. However, the detectability of the LLCP in our extended model is highly sensitive to the interplay of many-body terms and may be masked by solidification or appear as a broad crossover.

Specifically, we demonstrated that the balance among many-body components of the HB interaction can stabilize the low-density and high-density structural changes associated with the liquid-liquid phase transition. Interestingly, because an excessively strong three-body component of the HB interaction tends to promote ice formation, the lack of competition among various many-body contributions explains why certain models do not exhibit polyamorphism \cite{Moore2009, Chandler2011}. Consequently, understanding the nature of the MBIs in water, particularly nanoconfined water, requires further experimental and theoretical investigation.


\section*{ACKNOWLEDGMENTS}

We acknowledge the Grants n. 
PID2021-124297NB-C31 and PID2024-157478NB-C31 
funded by MCIN/AEI/10.13039/501100011033 and ERDF, EU.

\section*{AUTHOR DECLARATIONS}

\subsection*{Conclift of Interest}

The authors have no conflicts to disclose.


\bibliography{hexatic}

\clearpage

\begin{center}
\Large{Supplementary Information}
\end{center}

\section{Simulation Details}
\label{SimulationDetails}

\subsection{Metropolis Monte Carlo}
\label{MC}

We perform Monte Carlo (MC) simulations with a fixed number of molecules, $N=30^2=900$, at constant pressure $P$ and temperature $T$, while allowing the volume $V$ to fluctuate. The Metropolis algorithm is employed to simulate the system by randomly selecting one of the $5N+1$ variables--specifically, the $N$ coordinate vectors $r_i$, the $4N$ bonding indices $\sigma_{ij}$, and the volume $V_0$--and attempting to assign it a new random value. One MC step updates all $5N+1$ variables in a random sequence.

To reduce update rejections, we use an adaptive step-size algorithm. As a result, the parameters $\delta r$ and $\delta V$, defined in the following, are dynamically optimized to maintain a target acceptance ratio of $\approx 40\%$ \cite{Talbot2003, Bouzida1992}. 

{\bf Bonding indices update.} For each molecule $i$, we randomly select one of the four possible bonding indices $\sigma_{ij}$  facing a molecule $j$, then assign a new value $\sigma_{ij} \in \{1,2, \dots, q\}$ different from the previous value. 
The acceptance probability for the new configuration is $\exp [-\beta \Delta H]$ if $\Delta H > 0$ and  1 otherwise, where $\beta\equiv 1/k_BT$. 
Here, $\Delta H \equiv \Delta U + P \Delta V$ is the enthalpy change due to the formation (or breaking) of a HB, which affects the internal energy $U \equiv \langle \mathcal{H} \rangle$ and the system's volume change $\Delta V = v_{\mathrm{HB}} \Delta N_{\mathrm{HB}}$, where $\Delta N_{\mathrm{HB}}$ is the cange in the number of HBs. 
This MC update is the same as that performed for the original FS model.

{\bf System volume update.} We set $V_0^{\text{new}} = V_0^{\text{old}} + \varepsilon_{V}$, where $\varepsilon_{V} \in [-\delta V, + \delta V]$. 
Since the system may exhibit different symmetries, we allow changes in the aspect ratio of the system by modifying one of the two possible sides of the simulation box, $L_{x}$ and $L_{y}$, chosen at random with equal probability at every volume-change attempt by $\varepsilon_{V}$.
We accept this change with probability $(V_{0}^{\text{new}}/V_{0}^{\text{old}})^{N} \mathrm{exp}[-\beta \Delta H]$ if $\Delta H - Nk_BT \log (V_{0}^{\text{new}}/V_{0}^{\text{old}}) > 0$, and with probability 1 otherwise \cite{NPTMC}. 
This MC update differs from that performed for the original FS model, where the $L_{x}$ and $L_{y}$ symmetry was always preserved.

{\bf Coordinates update.} We perform a trial displacement of a randomly selected molecule, $r_{\alpha,\text{new}} = r_{\alpha,\text{old}} + \varepsilon_r$, where $\varepsilon_r \in [-\delta r,+\delta r]$ is a uniform random number and $\alpha=\{x,y\}$ indicates the $x$ or $y$ coordinate, while the new coordinates are constrained to lie within the cell of the selected molecule. 
We accept the new configuration with probability $\exp [-\beta \Delta U]$ if the change in internal energy $\Delta U > 0$, and with probability 1 otherwise. 
Note that in this type of update, there is no change in volume, nor any breaking or formation of HBs. The coordinate update is not present in the original FS model, in which the molecular coordinates are coarse-grained at the cell size.

Constraining the coordinate update of each molecule within its own cell imposes a strong condition of homogeneity that may lead to artifacts in a structural study.
To overcome this limitation and release this constraint, at the end of each MC step (update of all $5N+1$ system's variables in a random sequence), we perform $\mathcal{M}$ update of the coordinates, keeping the remaining degrees of freedom fixed.
We use $\mathcal{M}=10^{4}$, or until a molecule displaces more than $2l$ from its original position, where $l\equiv\sqrt{V/N}$ is the lateral size of a cell.
These additional $\mathcal{M}$ steps allow the coordinates to relax, losing memory of the underlying cell partition, and to optimize the molecular positions.

\subsection{Simulation protocol}
\label{SP}

To study the low-pressure, low-temperature region ($P<1$ and $T<0.1$) of the phase diagram, we start from a perfect square crystal at density $\rho=0.4$ with a fully formed HB network, with all bond indices $\sigma_{ij}$ in the same state. 
This situation corresponds to a configuration very close to the system's ground state at very low $T$. We then simulate the system by increasing $T$ in small steps, keeping $P$ constant and allowing the system to equilibrate at each stage. We accumulate statistics at each stage for at least $10^{5}$ MC steps. We follow a chained protocol in which the starting configuration at each new $T$ is the last configuration of the previous simulation.

On the other hand, in the high-pressure region ($P>1$), the energy is optimized for HD configurations, where we expect very few HBs to form, resulting in a close-packed order.
Therefore, we start these series of simulations from a perfect triangular crystal with no HBs and random bond states, which is much closer to the ground state energy under these conditions. 
In this case, we slowly decrease $P$ while keeping $T$ constant, following a chained protocol.

\section{Spatial correlations}
\label{S-Corr}

\subsection{Order Parameters}

In this section, we introduce several order parameters and analyze them to thoroughly characterize the system's structure. The local bond-orientational order parameter $\phi$ and the local translational order parameter $\psi$ are the two most significant parameters for this purpose. These quantities are particularly useful because they tend to vanish as the system approaches disordered phases, whereas they reach maxima as the system becomes more ordered.

The local bond-orientational order parameter associated with square symmetry is the 4-fold 
\begin{equation}
\phi_{4,j} \equiv \frac{1}{N_{\mathrm{nn}}} \sum^{N_{\mathrm{nn}}}_{k} \exp[i4\theta_{jk}],
\label{localphi}
\end{equation}
where $\theta_{jk}$ is the angle between the line connecting molecule $j$ to one of its nearest-neighbor molecules $k$ and an arbitrary axis, with the number of nearest neighbors fixed at $N_{\mathrm{nn}} = 4$. 
The corresponding global bond-orientational structural order parameter is:
\begin{equation}
	\Phi_4 \equiv \frac{1}{N} \sum^N_{j} \phi_{4,j} 
\end{equation}

For the translational order, the local and global order parameters associated with square symmetry are, respectively:
\begin{equation}
\psi_{\vec{G}_{\mathrm{Sqr}} j} \equiv
\exp[i \vec{G}_{\mathrm{Sqr}} \cdot \vec{r}_{j}],
\end{equation}
and 
\begin{equation}
	\Psi_{\vec{G}_{\mathrm{Sqr}}} \equiv \frac{1}{N} \sum_{j}^{N} \psi_{\vec{G}_{\mathrm{Sqr}} j} ~,
\end{equation}
where $\vec{r}_{j}$ are the coordinates of molecule $j$ and $\vec{G}_\mathrm{Sqr}$ is a vector in the reciprocal lattice of a crystal with square symmetry that best fits the current configuration of the system. 
For instance, in a system exhibiting global order aligned with the crystal's principal axes, $\vec{G}_{\mathrm{Sqr}} \equiv 2\pi/a\ \hat{e}_{x}$, where $a\equiv L_{x}/\sqrt{N}$ is the lattice constant, $L_x$ is the lateral system size such that $V_{0}=L_{x} L_{y}$ is the system's volume, and $\hat{e}_{x}$ is the unit vector along the $x$ axis.

The order parameters with triangular symmetry share analogous definitions. The 6-fold bond-orientational order parameter is:
\begin{equation}
	\Phi_6 \equiv \frac{1}{N} \sum^N_{j} \phi_{6,j} \equiv \frac{1}{N} \sum^N_{j} \frac{1}{N_{\mathrm{nn}}} \sum^{N_{\mathrm{nn}}}_{k} \exp[i6\theta_{jk}] 
\end{equation}
choosing a number of nearest neighbors $N_{\mathrm{nn}} = 6$. The translational order parameter is:
\begin{equation} 
	\Psi_{\vec{G}_{\mathrm{Tri}}} \equiv \frac{1}{N} \sum_{j}^{N} \psi_{\vec{G}_{\mathrm{Tri}},j} \equiv \frac{1}{N} \sum_{j}^{N} \exp[i \vec{G}_{\mathrm{Tri}} \cdot \vec{r}_{j}] 
\end{equation}
where we choose $\vec{G}_{\mathrm{Tri}} \equiv 4\pi / \sqrt{3} a\ \hat{e}_{y}$ if the lateral size of the system $L_{x} > L_{y}$, or $\vec{G}_{\mathrm{Tri}} \equiv 4\pi / \sqrt{3} a\ \hat{e}_{x}$ otherwise.

\subsection{The hexatic phase}

\captionsetup[subfigure]{position=top, labelfont=bf,textfont=normalfont,singlelinecheck=off,justification=raggedright}

\begin{figure}
	\includegraphics[width=0.45\linewidth]{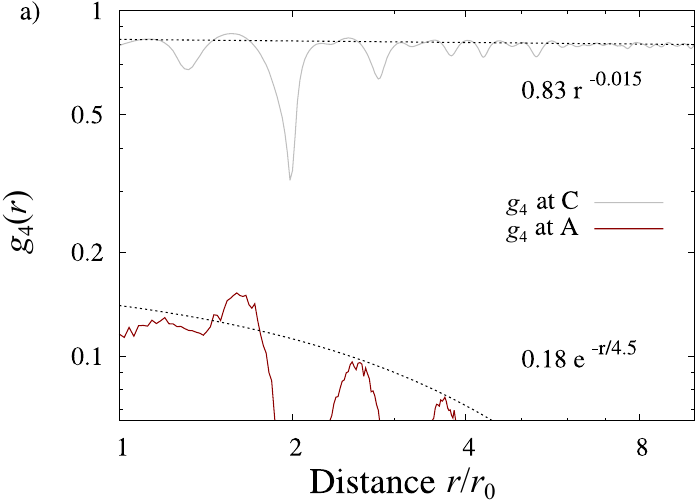}
	\includegraphics[width=0.45\linewidth]{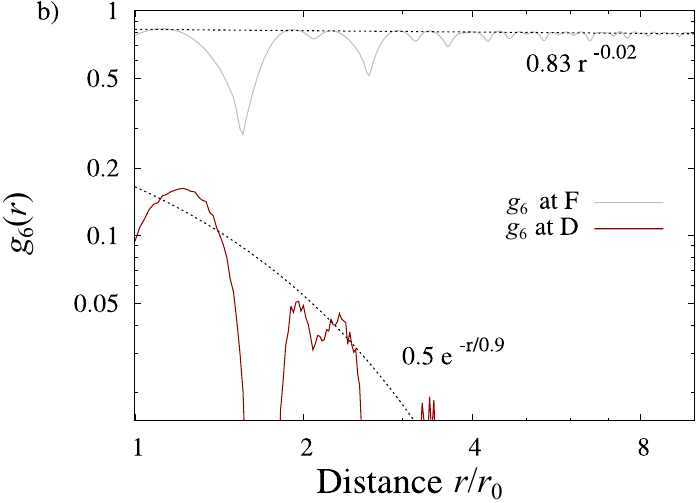}\\
	\includegraphics[width=0.45\linewidth]{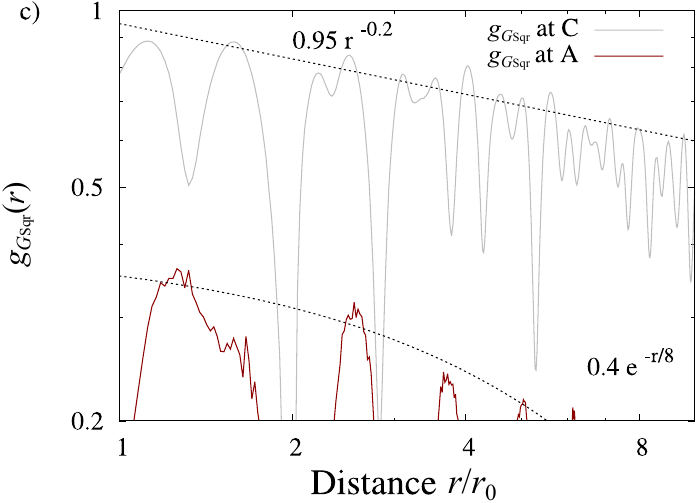}
	\includegraphics[width=0.45\linewidth]{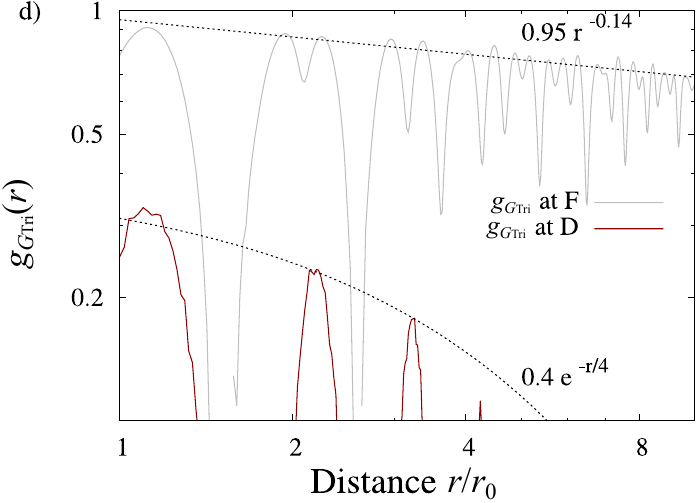}
	
	\caption{Correlation functions $g_{\alpha}(r)$ and $g_{\vec{G}}(r)$ (solid lines) and their corresponding fitting functions (dashed lines), computed at different state points. 
	The state point labels refer to those indicated in Fig. 1a of the main text: 
	A at $(Tk_B/4\epsilon=0.15, Pv_0/4\epsilon=0.05)$,
	C at $(Tk_B/4\epsilon=0.06, Pv_0/4\epsilon=0.05)$,
	D at $(Tk_B/4\epsilon=0.18,Pv_0/4\epsilon=0.2)$,
	F at $(Tk_B/4\epsilon=0.12, Pv_0/4\epsilon=0.8)$.
	{\bf a)} 4-fold bond-orientational correlation function in A (red) 
	and C (gray); 
	{\bf b)} 6-fold bond-orientational correlation function in D (red) 
	and F (gray); 
	{\bf c)} translational square correlation function in A (red) 
	and C (gray); 
	{\bf d)} translational triangular correlation function in D (red) 
	and F (gray). }
	\label{fig:OrderParamsCorrelation}
\end{figure}

We study the spatial correlations of the local order parameters, $\phi$ and $\psi$, introduced in the previous section. We retain the same nomenclature as for the radial distribution function $g_\alpha(r)$:
\begin{align}
	g_4(r) &\equiv \operatorname{Re} [ \langle \phi^{*}_{4}(0) \phi_{4}(r) \rangle ]\\
	g_6(r) &\equiv  \operatorname{Re} [ \langle \phi^{*}_{6}(0) \phi_{6}(r) \rangle ] \\
	g_{\vec{G}_{\mathrm{Sqr}}}(r) &\equiv  \operatorname{Re} [ \langle \psi^{*}_{\vec{G}_{\mathrm{Sqr}}}(0) \psi_{\vec{G}_{\mathrm{Sqr}}}(r) \rangle ]\\
	g_{\vec{G}_{\mathrm{Tri}}}(r) &\equiv  \operatorname{Re} [ \langle \psi^{*}_{\vec{G}_{\mathrm{Tri}}}(0) \psi_{\vec{G}_{\mathrm{Tri}}}(r) \rangle ]
\end{align} 
where $\phi_4(0)$ is defined by Eq.(\ref{localphi}) for molecule $i$ at the origin, $r=r_{ij}$ is the distance between molecules $i$ and $j$,
$\phi_4^{*}$ is the complex conjugate of $\phi_4$, and analogous definitions apply to the other functions in $g_\alpha(r)$. 
We average over all molecule pairs $(i,j)$, and over $N_{t}$ independent configurations, i.e., $\left< \phi(0)\phi(r)\right> \equiv N_{t}^{-1}\sum_{t} [N(N-1)/2]^{-1}\sum_{(i,j)} \phi(\vec{r}_{i};t)\phi(\vec{r}_{j};t)$. 
Since these are complex-valued functions, we take the real part, Re[$\cdot$], of these quantities for convenience.

To characterize the various phases of the system, we study the functional forms of the correlations in each region of the phase diagram (Fig. \ref{fig:OrderParamsCorrelation}). 
In liquids (Table \ref{Table2}), where long-range order is nonexistent by definition, all spatial correlations, both orientational and translational, are short-ranged and decay exponentially, i.e., $g_\alpha(r) \sim \exp (-r/\xi)$ (with $\alpha=\{4,6\}$) and $g_{\vec{G}} \sim \exp (-r/\xi)$. 
On the other hand, a 2D crystal exhibits long-range orientational order $g_{\alpha}(r) \sim \mathrm{constant}$, but quasi-long-range translational order with geometric decay of the correlations $g_{\vec{G}} \sim r^{-\eta}$. 
The hexatic phase is an intermediate state in which the liquid transforms continuously into a solid. This phase exhibits quasi-long-range orientational order $g_{\alpha}(r) \sim r^{-\eta}$ and short-range translational order $g_{\vec{G}} \sim \exp (-r/\xi)$. Some examples of these functions, computed in different regions of the phase diagram, and their corresponding functional fits are shown in (Fig. \ref{fig:OrderParamsCorrelation}).

\begin{table}
\begin{tabular}{ |c||c|c| } 
 \hline
 Phase & Orientational correlation & Translational correlation\\
 \hline
 \hline
 Liquid & $g_\alpha(r) \sim \exp (-r/\xi)$ & $g_{\vec{G}} \sim \exp (-r/\xi)$ \\ 
 \hline
Hexatic & $g_{\alpha}(r) \sim r^{-\eta}$ & $g_{\vec{G}} \sim \exp (-r/\xi)$\\
 \hline
  Crystal& $g_{\alpha}(r) \sim \mathrm{constant}$ & $g_{\vec{G}} \sim r^{-\eta}$\\ 
 \hline
\end{tabular}
\caption{Summary of spatial correlations decay for different phases in 2D, with $\alpha=\{4,6\}$ for square (LD) and triangular (HD) symmetries.  Exponential decay corresponds to short-ranged order, power-law decay to quasi-long-range order, and constant behavior to long-range order.
}
\label{Table2}
\end{table}

\subsection{The phases limits}

To study the limits of the LD and HD crystal (ice) phases relative to their corresponding hexatic phases, we fit the translational correlation functions at any state point to an exponential law $g_{\vec{G}_\alpha} (r) = C \exp(-r/\xi)$, fixing a characteristic length $\xi=L/2$ and using the pre-factor $C$ as the single fitting parameter, and calculate the corresponding coefficient of determination $R^{2}$.
The coefficient $R^{2}$ measures a model's goodness of fit. In regression analysis, $R^{2}$ quantifies the proportion of variance in the response variable explained by the predictor variables. An $R^{2}$ value of 1 indicates a perfect fit between the model and the data, whereas 0 indicates the model explains no variability.

\begin{figure*}
\subfloat{
\includegraphics[width=0.4\linewidth]{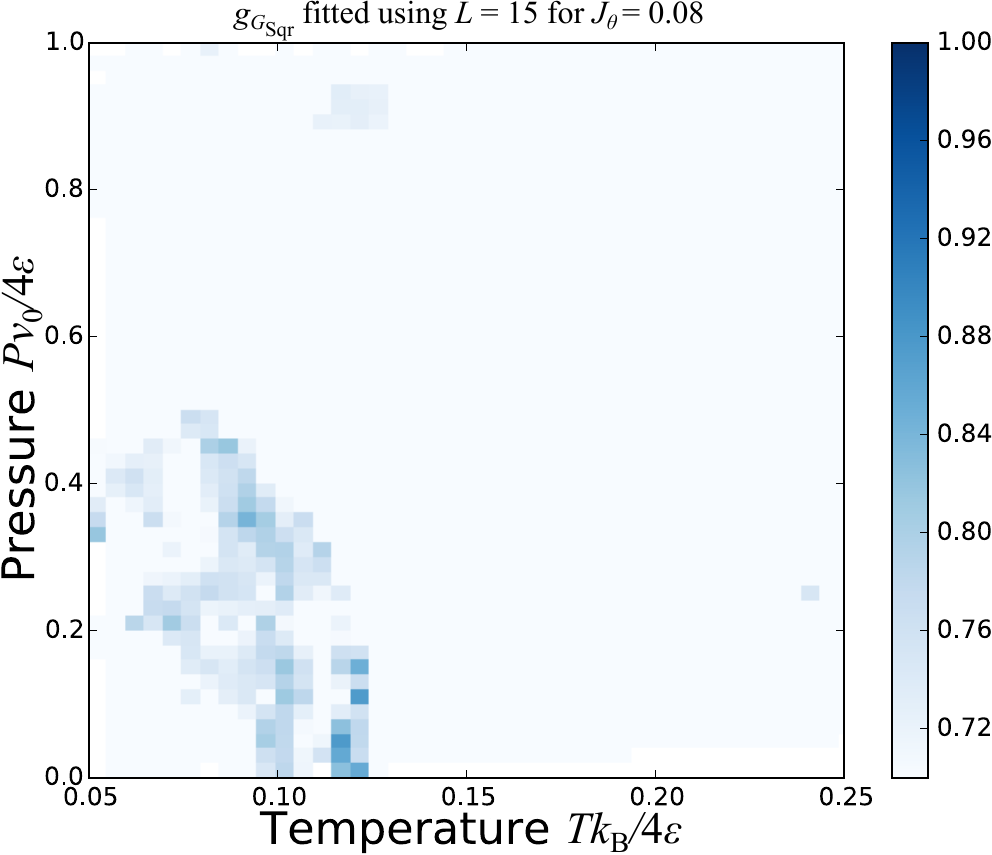}
}
\subfloat{
\includegraphics[width=0.4\linewidth]{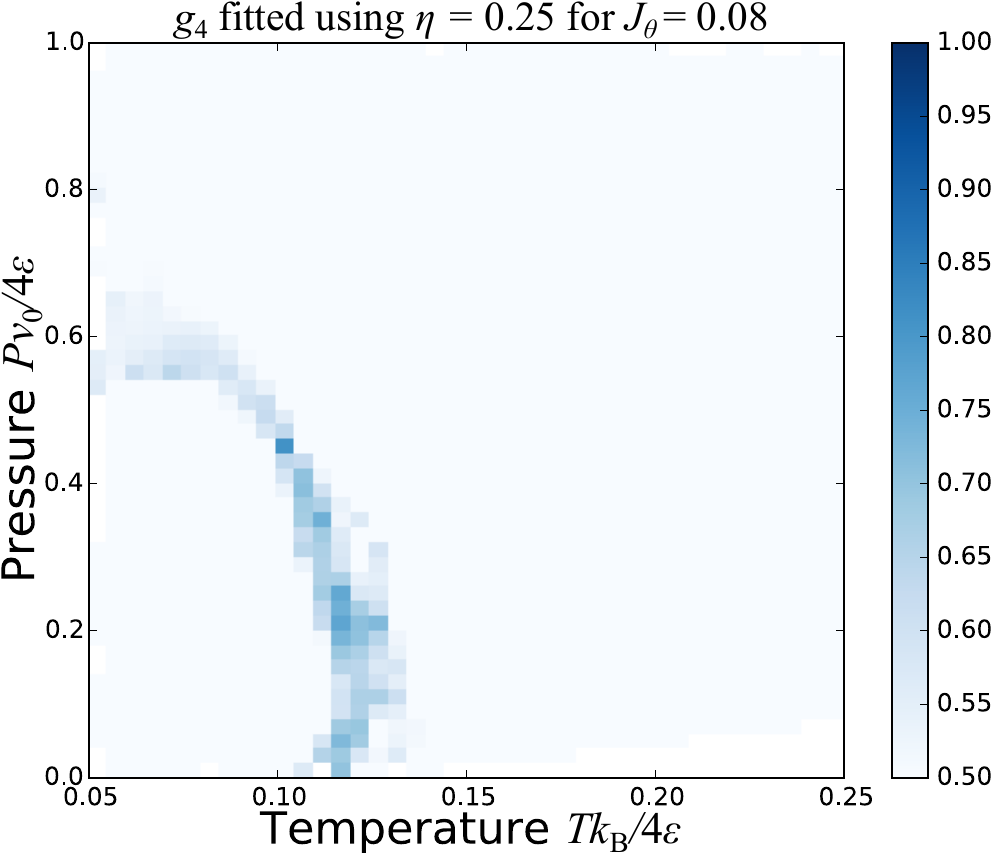}
}\\
\subfloat{
\includegraphics[width=0.4\linewidth]{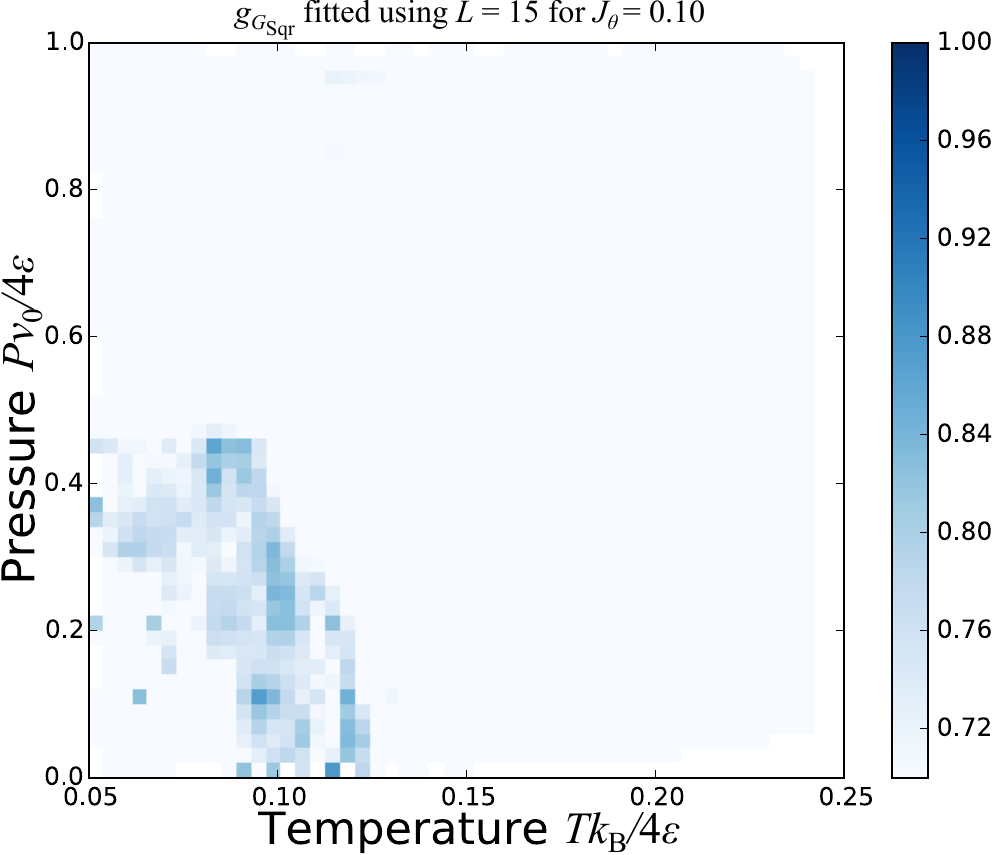}
}
\subfloat{
\includegraphics[width=0.4\linewidth]{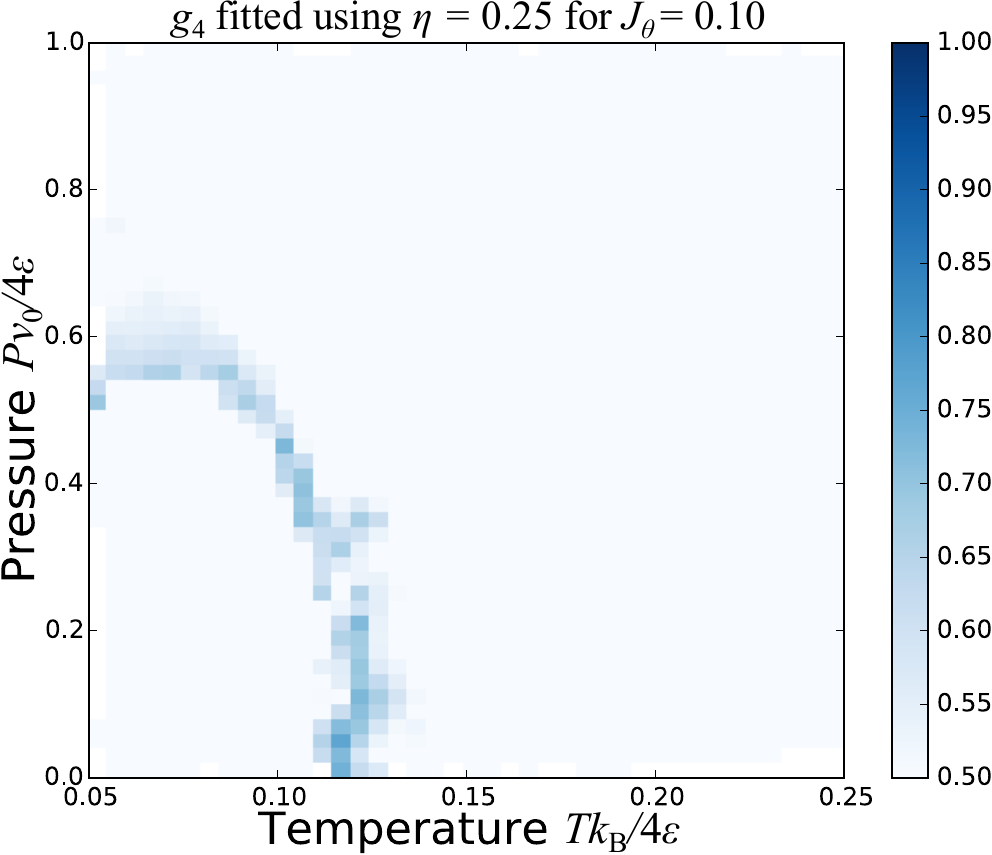}
}\\
\subfloat{
\includegraphics[width=0.4\linewidth]{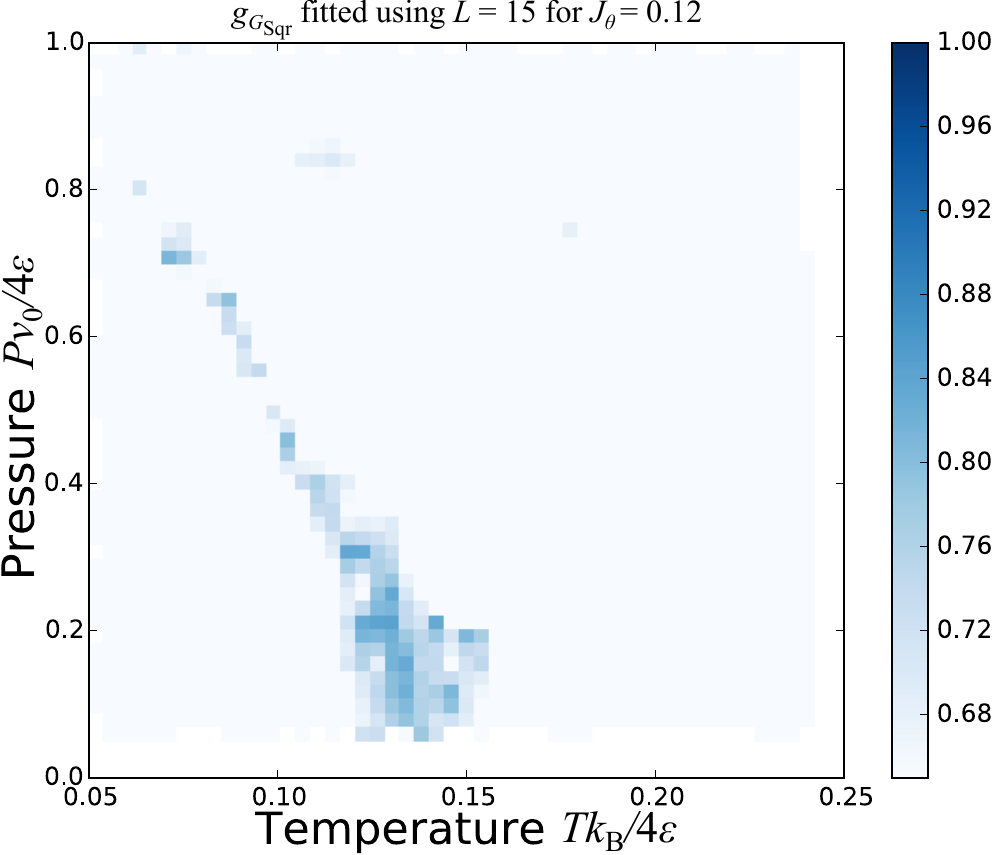}
}
\subfloat{
\includegraphics[width=0.4\linewidth]{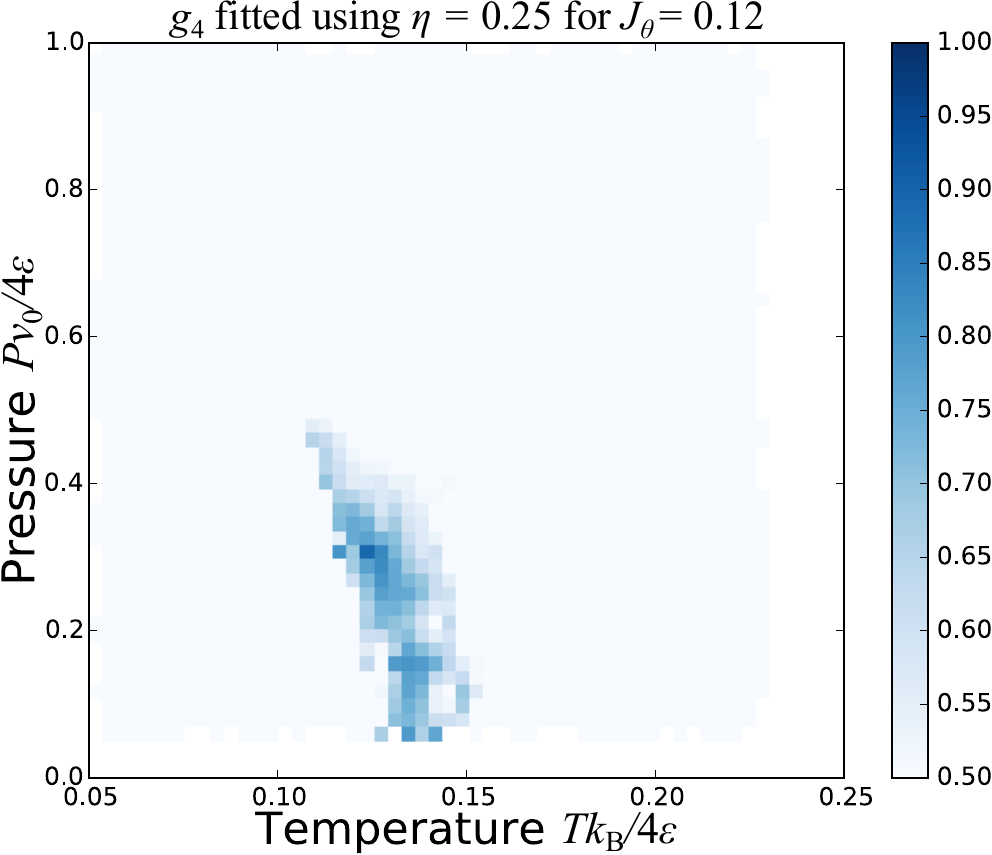}
}
\caption{Goodness of fit (coefficient of determination $R^{2}$) of the translational correlation function (left) and the orientational correlation function (right) to the exponential decay model at different state points $(T,P)$ at low density (square symmetry). $T$ is measured in $4\epsilon/k_B$, while $P$ is in $4\epsilon/v_0$. The color code on the right quantifies the value of $R^{2}$. In each column, the panels are for $J_{\theta}/4\epsilon = 0.08$, 0.10, 0.12 from top to bottom. The other model parameters are indicated in Table 1 of the main text.} 
\label{fig:FittingSquare}
\end{figure*}

\begin{figure*}
\subfloat{
\includegraphics[width=0.4\linewidth]{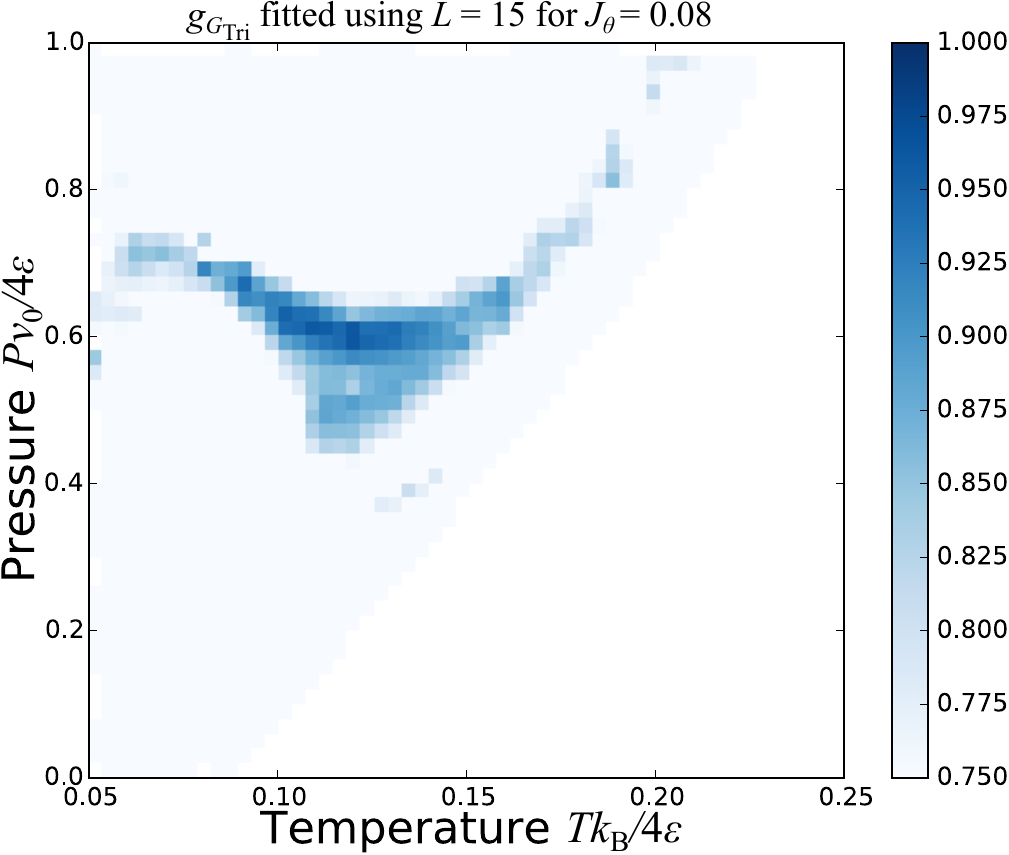}
}
\subfloat{
\includegraphics[width=0.4\linewidth]{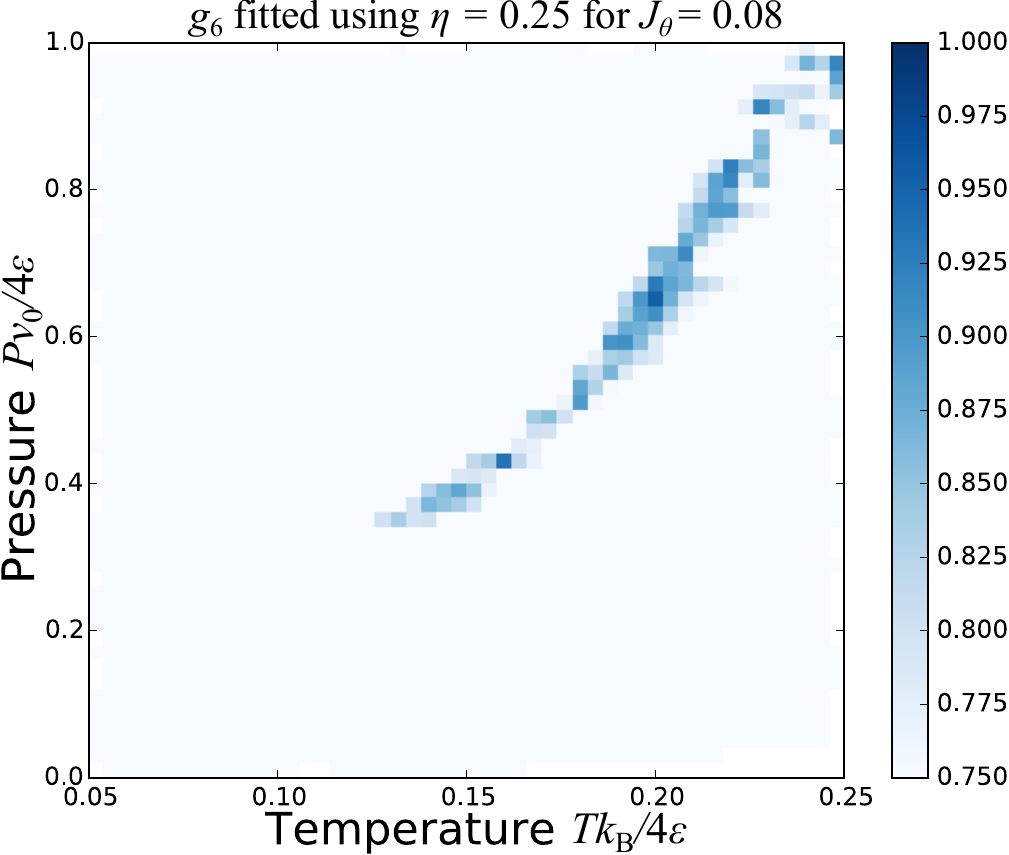}
}\\
\subfloat{
\includegraphics[width=0.4\linewidth]{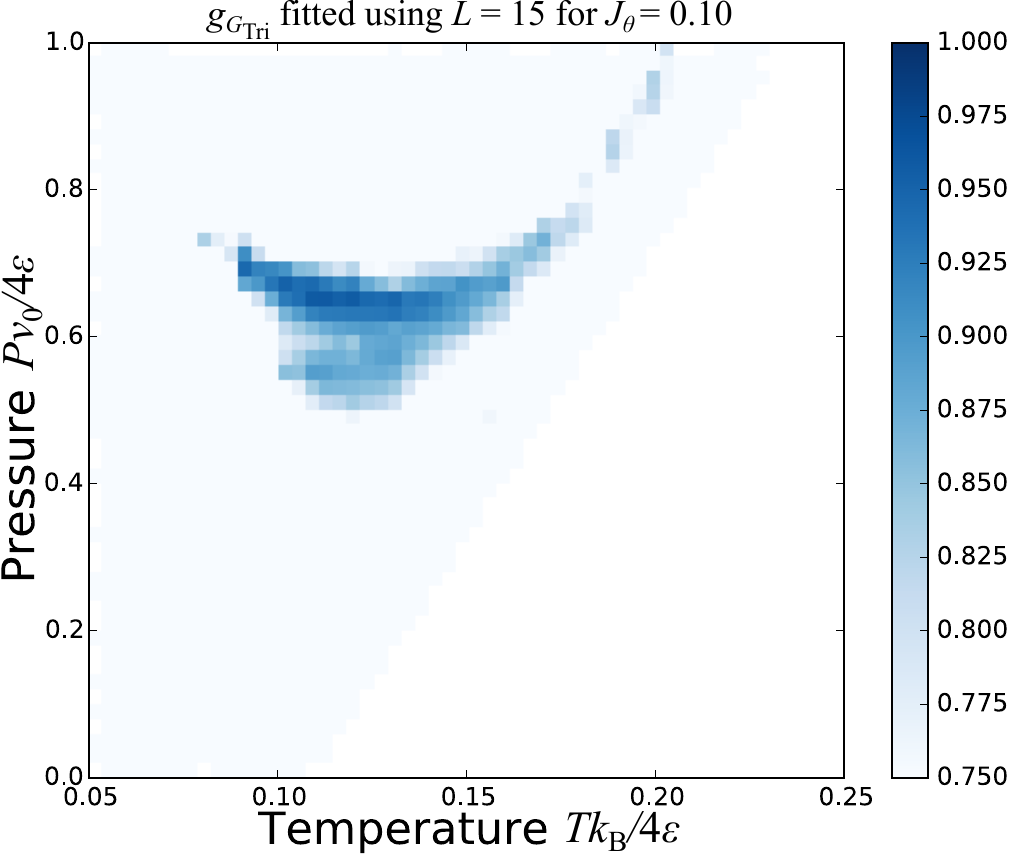}
}
\subfloat{
\includegraphics[width=0.4\linewidth]{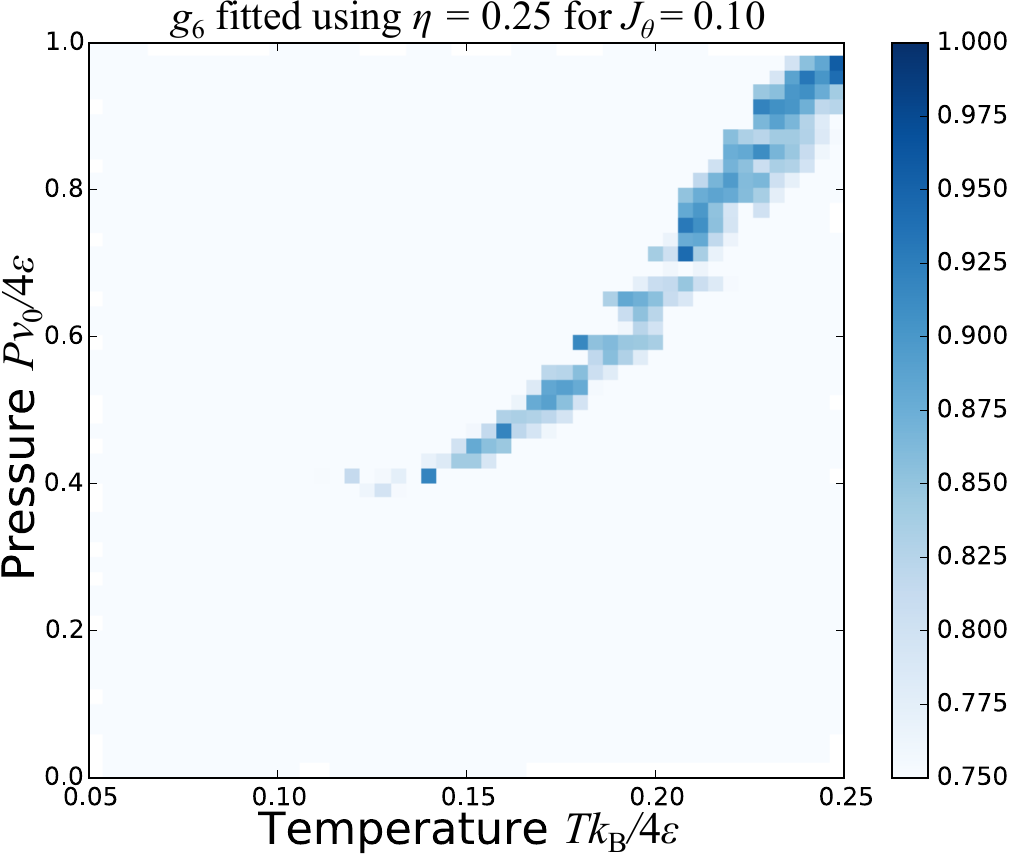}
}\\
\subfloat{
\includegraphics[width=0.4\linewidth]{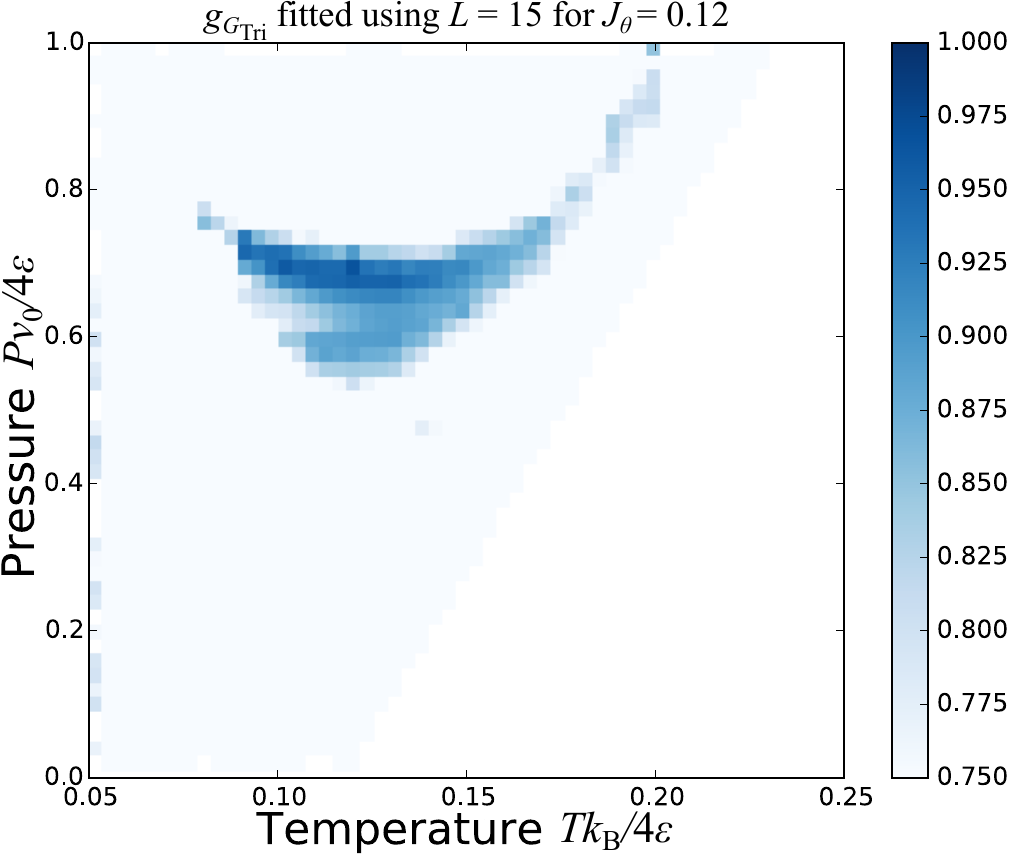}
}
\subfloat{
\includegraphics[width=0.4\linewidth]{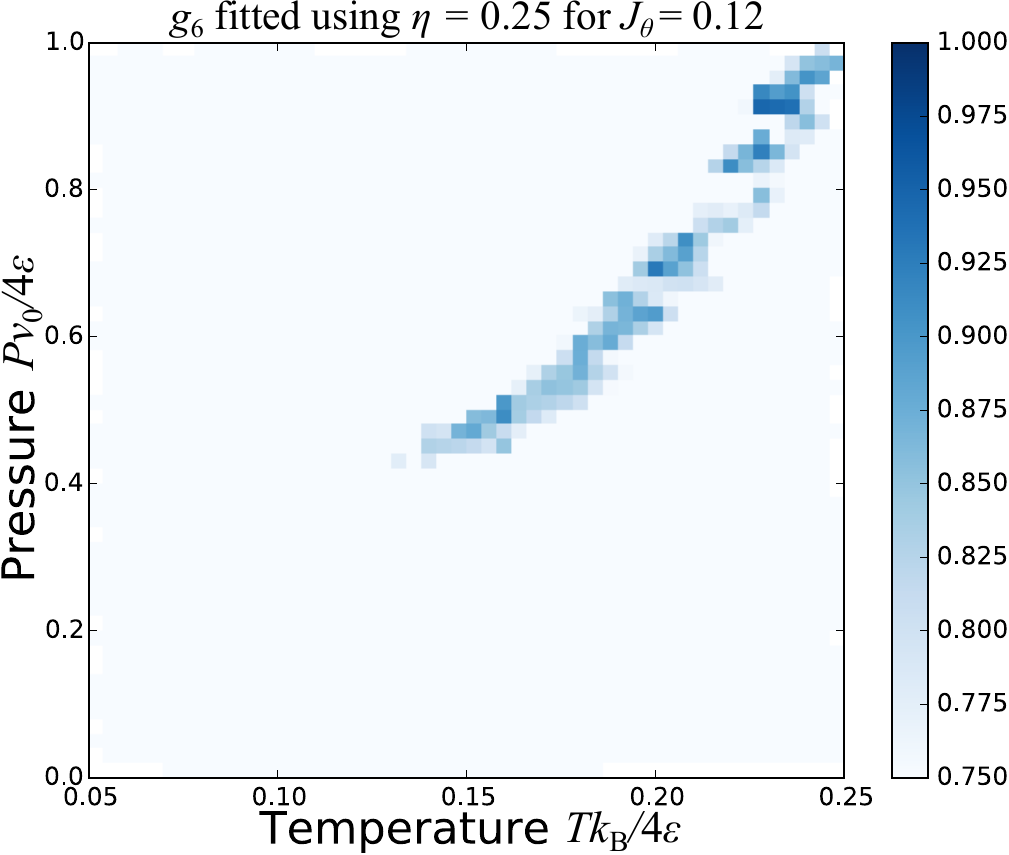}
}
\caption{As in Fig. \ref{fig:FittingSquare} but at high density (triangular symmetry).} 
\label{fig:FittingTriangular}
\end{figure*}

As expected, $R^{2}$ reaches a maximum along the melting 
line, where correlations between molecular coordinates become as large as the system size. 
The LDi-LDh line can be observed in the left column of Fig. \ref{fig:FittingSquare} by fitting $g_{\vec{G}_{\rm{Sqr}}}(r)$, while the HDi-HDh line appears in the left column of Fig. \ref{fig:FittingTriangular} by fitting $g_{\vec{G}_{\rm{Tri}}}(r)$.

Analogously, we study the limit of the liquid phases relative to the hexatic phases by fitting the orientational order parameter correlations with a power-law function $g_{\alpha}(r) = C r^{-\eta}$. 
As predicted by the KTHNY theory, the onset of the hexatic phase occurs when the correlation function decays with $\eta=0.25$ \cite{Kosterlitz1972, Kosterlitz1973, Nelson1977, Halperin1978, Young1979}.
We apply the same procedure described in the previous analysis to assess the goodness of fit across the phase diagram. 
The LDL-LDh line appears in the right column of Fig. \ref{fig:FittingSquare} after fitting $g_{4}(r)$, while the HDL-HDh line appears in the right column of Fig. \ref{fig:FittingTriangular} after fitting $g_{6}(r)$.
In all cases, we find that the best-fit lines lie at higher $T$ than the solid-hexatic transition lines. 
This allows us to delimit the extent of the hexatic phases and to determine the stability limits of the liquid and solid phases.

As a demonstration of our analysis's effectiveness, we observe that the authors of Ref.\cite{Hoang2016} reported RDFs resembling those in Fig. 2 of the main text. They attributed these to a single first-order phase transition between solid and liquid, while our detailed analysis distinctly identifies those RDFs with the liquid, hexatic, and solid phases.

\end{document}